\documentclass[aps,prb,twocolumn]{revtex4-1}

\usepackage{graphicx}
\usepackage{amsmath}
\usepackage{amsfonts}
\usepackage{bm}
\usepackage{mathtools}
\usepackage{tikz}

\def\BFA{{BF approximation}}
\def\qR{{\bf R}}                                   
\def\qr{{\bf r}}                                   
\def\qk{{\bf k}}                                   
\def\qq{{\bf q}}                                   

\renewcommand{\Re}{\operatorname{\mathfrak{Re}}}
\renewcommand{\Im}{\operatorname{\mathfrak{Im}}}

\renewcommand{\vec}[1]{\bm{#1}}

\newcommand{\diff}[1]{\mathrm{d} #1}

\newcommand{\order}[1]{\mathcal{O}(#1)}

\newcommand{\hamiltonian}{{\mathcal{H}}}
\newcommand{\rv}{{\qr}}
\newcommand{\Rv}{{\qR}}
\newcommand{\kv}{{\qk}}

\newcommand{\sumprime}{\sideset{}{'}\sum}
\newcommand{\deltaHamiltonian}{\delta \hamiltonian}
\newcommand{\potential}{V}
\newcommand{\potentialTilde}{\widetilde{\potential}}
\newcommand{\excitationOp}{\delta U}
\newcommand{\deltau}{\delta u}
\newcommand{\deltauTilde}{\delta \tilde{u}}
\newcommand{\deltauDot}{\delta \dot{u}}

\newcommand{\deltarho}{\delta \rho}
\newcommand{\deltarhoDot}{\delta \dot{\rho}}
\newcommand{\deltarhoTilde}{\delta \tilde{\rho}}
\newcommand{\deltaX}{\delta X}
\newcommand{\deltaXTilde}{\delta \widetilde{X}}
\newcommand{\deltaXTildeDot}{\delta \dot{\widetilde{X}}}

\DeclareMathOperator*{\sumint}{%
\mathchoice%
  {\ooalign{$\displaystyle\sum$\cr\hidewidth$\displaystyle\int$\hidewidth\cr}}
  {\ooalign{\raisebox{.14\height}{\scalebox{.7}{$\textstyle\sum$}}\cr\hidewidth$\textstyle\int$\hidewidth\cr}}
  {\ooalign{\raisebox{.2\height}{\scalebox{.6}{$\scriptstyle\sum$}}\cr$\scriptstyle\int$\cr}}
  {\ooalign{\raisebox{.2\height}{\scalebox{.6}{$\scriptstyle\sum$}}\cr$\scriptstyle\int$\cr}}
}

\begin{document}

\title{The Multi-component Correlated Basis Function Method and its Application to
Multilayered Dipolar Bose Gases}

\author{Michael Rader$^{1,2}$, Martin Hebenstreit$^{1,2}$, and Robert E. Zillich$^2$}

\affiliation{$^1$Institute for Theoretical Physics, University of Innsbruck,
Technikerstr. 21a, 6020 Innsbruck, Austria\\
$^2$Institute for Theoretical Physics, Johannes Kepler University, Altenbergerstrasse 69, 4040 Linz, Austria
}

\begin{abstract}
We present a method for calculating the dynamics of a bosonic mixture,
the multi-component correlated basis function (CBF) method.  For single components,
CBF results for the excitation energies agree quite well with experimental results,
even for highly correlated systems like $^4$He, and recent systematic improvements of CBF achieve perfect agreement.
We give a full derivation of multi-component CBF, and
apply the method to a dipolar Bose gas cut into two-dimensional layers
by a deep optical lattice, with coupling between layers due to the long-ranged dipole-dipole
interaction.  We consider the case of strong coupling, leading to large positive
interlayer correlations.  We calculate the spectrum for a system of 8 layers and
show that the strong coupling can lead to a simpler spectrum than in the uncoupled
case, with a single peak carrying most of the spectral weight.
\end{abstract}


\maketitle

\section{Introduction}

Many experiments in the field of ultracold quantum gases work with multi-component systems.
The components can be comprised of different atom species, of the same atoms but different
isotopes or in different hyperfine states, or of the same atoms but separated
spatially by an optical lattice deep enough to prevent tunneling.  If there were no
interactions between different components, the multi-component system would be
an ensemble of independent systems, one for each component.  This is the case for example
for a multi-layered Bose gas if the interaction is short-ranged and thus the interaction
between different layers is negligible.  Interaction between different components
can create a host of new phenomena.  The components may unmix
or conversely, they may create a liquid phase out of two gas-like components.
Polaron physics can be studied if one component is very dilute.
Eventually, the interaction between two different atoms can be tuned by a Feshbach
resonance to lead to the creation of weakly or deeply bound heteronuclear molecules.

In this paper we are interested in the dynamics of a coupled multi-component Bose system.
For this purpose we generalize the correlated basis function (CBF) method to homogeneous
systems of many species of bosons.  The CBF method was introduced by Jackson and
Feenberg.\cite{JaFe,JaFe2}  When one follows an alternative route to CBF, based on
linear response theory, see e.\,g.\ Refs.~\onlinecite{Saarela86,Clements93}, it becomes clear
that the CBF method belongs to a family
of methods based on a time-dependent variational ansatz of the many-body wave function
that accounts for correlations.  The area of application is the dynamics of
quantum many-body systems where the interaction has a dominant influence, as opposed
to systems amenable to mean field approaches.  The simplest member of this family is the
approximation by Bijl\cite{Bijl40} and Feynman\cite{Feynman3}, where the pair and higher correlations are still
assumed to be time-independent.  In the CBF method, time-dependent pair correlations
are taken into account, which in the present case of multi-component systems
depend also on the two components.

We apply the multicomponent CBF method to a homogeneous system
to keep the computational effort to a minimum.  A two-dimensional multi-component
Bose gas is realized by placing a Bose gas into a deep 1D optical lattice,
producing 2D layers.  Although the same kind of particles are loaded in each
layer, the spatial separation into layers makes particles in different layers distinguishable,
thus generating a multicomponent system.  In the absence of tunneling and long-ranged interactions,
this would be just an ensemble of uncoupled 2D systems.  Coupling can be achieved
by a long-range interaction, such as the dipole-dipole interaction, which is felt
both between particles in the same layer (intra-layer) and on different layers (inter-layer).
The dipole-dipole interaction has a range on the order of
the dipole length $r_D=mD^2/(4\pi\epsilon_0\hbar^2)$, where $D$ is the dipole
moment of a particle.\cite{baranovPhysRep08,baranovChemPhys12}
$r_D$ can well exceed the distance between neighboring layers, determined by
the wavelength of the laser for the optical trap.  This is especially true
if the particles are heterogeneous Bose condensed molecules with an electric
dipole moment, which are being studied experimentally.
\cite{sagePRL05,deiglmayrPRL08,ospelkausNatPhys08,niScience08,voigtPRL09,shumanNature10,niNature10,stuhlNature12,takekoshiPRL14,molonyPRL14,parkPRL15_NaK,guoPRL16}
We are particularly
interested in those cases where $r_D$ is so large that the system is nearly
unstable against the formation of bound states between dipoles in different layers.
We note that also a {\em low} density in a layer promotes {\em strong} interlayer
correlations.  In this work we do not consider the case of even stronger coupling
leading to bound states.\cite{wangPRL07,maciaPRA14,filinovPRA16}

In a previous work,\cite{hebenstreitPRA16} we presented energetic and structural results for the liquid-like ground state
of two 2D dipolar Bose gas layers (with anti-parallel polarization), using the multi-component
generalization of the hyper-netted chain Euler-Lagrange (HNC-EL)
method.\cite{Chuckmix,OldChuckMix,MIL80,chakrabortyPRB82,krotscheckPhysRep93,hebenstreitbachelor}  In Ref.~\onlinecite{hebenstreitPRA16} we
also discussed the speed of sound, i.\,e.\ the long wave-length limit of collective excitations,
for which we used the Bijl-Feynman (BF) approximation, which gives reasonable results in the
long wave-length limit.  However, the BF energies start to deviate from the exact excitation energies as
the wave number increases, and it also does not account for broadening due to coupling between
excitations.   While the \BFA\ only considers
mean-field fluctuations (on top of a correlated ground state), the CBF method also accounts for fluctuations
of pair correlations in an approximate way.  The CBF methods has long been tested on strongly
correlated systems like $^4$He, both homogeneous~\cite{Saarela86,ChangCamp76}
and inhomogeneous~\cite{clementsPRB96,krotscheckJCP01},
where a considerable improvement with respect to the
simple BF estimate is achieved.  Further improvement can be achieved
systematically~\cite{campbellPRB09,campbellJLTP10,campbellPRB15} by e.\,g.\ incorporating fluctuations
of triplet correlations; this has been demonstrated for the case of a homogeneous system of a single
species of bosons, namely superfluid bulk $^4$He.
The dynamic structure function of $^4$He, which is a very strongly correlated quantum fluid,
was calculated using this improved theory, and the results are essentially
identical to measurements using inelastic neutron scattering, throughout
the whole experimentally accessible range of momenta and energies.\cite{beauvoisPRB16}

In section~\ref{sec:theory}, we derive the multi-component
CBF method, where the bulk of derivation is delegated to the appendix.  As an application we
present the dynamic response of coupled 2D layers of dipolar bosons in section~\ref{sec:results}.

\section{Correlated Basis Function Theory}
\label{sec:theory}

We consider a mix of several interacting bosonic components, described by the many-body Hamiltonian
\begin{align}
  \hamiltonian_0 = - \sum\limits_{\alpha} \sum\limits_{j}^{N_\alpha} \frac{\hbar^2}{2 m_\alpha} \nabla^2_{\alpha,j} 
  + V(\Rv)
\end{align}
where $\alpha$ denotes a component and $j$ enumerates the $N_\alpha$ particles within the component $\alpha$.
$\rv_{\alpha,j}$ are the coordinates of the particles and $\Rv$ is a shorthand notation for the entirety of particle coordinates.
$V$ denotes a pair-wise interaction potential,
\begin{align}
  V(\Rv) =  \frac{1}{2} \sumprime\limits_{\alpha, \beta, j, k}
      v_{\alpha \beta}(\rv_{\alpha,j},\rv_{\beta,k})\,.
\end{align}
The prime of the sum tells us to leave out the term with $\alpha = \beta$ and $j = k$.

We consider homogeneous 2D layers of Bose gas, where adjacent layers are seperated by a distance $d$ and
$\rho_\alpha$ is the partial density of layer $\alpha$; the total density is $\rho_{\rm tot}=\sum_\alpha\rho_\alpha$.
The particles in all layers shall be dipoles of the same species, with equal
masses $m$ and dipole moments $D$.  We follow the custom in this field and give length in units of
the dipole length $r_D=mD^2/(4\pi\epsilon_0\hbar^2)$, mentioned above, and energy in units of
$E_D=\hbar^2/(mr_D^2)$.  In dipole units, the kinetic energy operator becomes
$-{1\over 2}\sum_{\alpha} \sum_{j} \nabla^2_{\alpha,j}$.
The dipole moments are oriented perpendicular to the layer planes, leading to
an interaction potential (in dipole units) between a dipole in layer $\alpha$ and another dipole on layer $\beta$,
separated by a distance $z=d|\alpha-\beta|$,
\begin{align}
v_{\alpha \beta}(\rv,\rv') \equiv v_{\alpha \beta}(|\rv-\rv'|) = {|\rv-\rv'|^2 - 2 z^2 \over (|\rv-\rv'|^2 + z^2)^{5/2}}
\end{align}
where $\rv$ and $\rv'$ are the respective two-dimensional coordinates of the two dipoles with their
respective layer.

In this paper we are mainly interested in the dynamics of this multi-component Bose gas.  However,
before we can calculate dynamic properties, we need to obtain ground state properties which serve
as ingredients for the CBF method as shown in the derivation below.  The ground state can be
obtained, for example, from quantum Monte Carlo simulations.  This provides exact
ground state properties of dipolar quantum gases
\cite{astraPRL07,buechlerPRL07,maciaPRA11,maciaPRL12,maciaPRA14,hebenstreitPRA16,filinovPRA16},
but it would be computationally very demanding for the large number of layers that we study in this work.
For the fairly low partial denstities $\rho_\alpha$ that we consider here, it is sufficient to
use approximate methods.  The variational HNC-EL method is based on the
Jastrow-Feenberg ansatz~\cite{FeenbergBook} consisting of a product of pair correlation functions,
\begin{equation}
  \Psi_0 = \exp\Big[{1\over 4}\sideset{}{'}\sum_{\alpha,\beta,j,k}
               u_{\alpha\beta}(|\qr_{j,\alpha} - \qr_{k,\beta}|)\Big]\,.
\label{eq:Psi0}
\end{equation}
For a short discussion of this ansatz and further approximations, see Ref.~\onlinecite{hebenstreitPRA16}
where we used this ansatz for a dipolar bilayer.  For the dynamics we need structural
quantities like the pair distribution function
\begin{equation}
  g_{\alpha\beta}(|\qr_{\alpha} - \qr_{\beta}|)
= {N_\alpha(N_\beta-\delta_{\alpha\beta}) \over \rho_\alpha\rho_\beta} \int\! \diff{\tau}_{\alpha\beta}|\Psi_0(\Rv)|^2
\label{eq:g2}
\end{equation}
where the integral is over all particles except one in layer $\alpha$ and one in layer
$\beta$ (the notation $\diff{\tau}_{\alpha\beta}$ is defined in the beginning of the appendix).
$u_{\alpha\beta}$, and thus $g_{\alpha\beta}$ are determined using the Rayleigh-Ritz variational principle
which states that the ground state minimizes the energy: $E=\langle\Psi|H|\Psi\rangle$ is minimal for
the ground state $\Psi=\Psi_0$.  The resulting Euler-Lagrange equations $\delta E=0$, using the hypernetted-chain
approximation, are the HNC-EL equations.

We derive the multi-component CBF method via the linear response
approach, which has the benefit that we can relate the dynamics of the system directly to physical realizations
of perturbations.  The perturbated Hamiltonian is
\begin{align}
  \hamiltonian = \hamiltonian_0 + \deltaHamiltonian(t)
\end{align}
with a small time-dependent one-body potential as perturbation
\begin{align}
  \delta \hamiltonian = \sum_{\alpha} \sum\limits_{j}^{N_\alpha} \potential_{\alpha}(\rv_{\alpha,j}, t)
  \text{.}
\end{align}
The time-dependent perturbation potential $\potential_\beta$ can be different for each component.

Similarly to the Jastrow-Feenberg ansatz for the many-body ground state we introduce
an ansatz for the excited wave function
\begin{align}
  \Psi(\Rv, t)
  \propto
  e^{-i E_0 t} e^{\frac{1}{2} \excitationOp(\Rv, t)} \Psi_0(\Rv)
  \text{,}
\end{align}
assuming the ground state wave function $\psi_0$ with an energy $E_0$ is known.
The excitation operator is
\begin{align}
  \excitationOp = \sum\limits_{\alpha, j} \deltau_\alpha(\rv_{\alpha,j}) 
    + \frac{1}{2} \sumprime\limits_{\alpha, \beta, j, k}
      \deltau_{\alpha \beta}(\rv_{\alpha,j}, \rv_{\beta,k}, t)
\label{eq:deltaU}
\end{align}
with the one- and two-body correlation fluctuations $\deltau_\alpha$ and $\deltau_{\alpha \beta}$.
One-body ``correlation'' fluctuations $\deltau_\alpha$ can be considered as time-dependent mean-field approximation,
on top of a fully correlated ground state.  Neglecting two-body correlation fluctuations leads to the
multi-component generalization of the \BFA .  Two-body correlation fluctuations
are the first step beyond the \BFA, which improves the accuracy of the
excitation energy and accounts for damping due to coupling between excitations, but at the price
of technically more complex derivation, which we therefore relegate to the appendix.  The inclusion of
three-body correlations has been achieved only for a single-component Bose system
and impurities therein~\cite{campbellJLTP10,campbellPRB15}
and applied with great success to superfluid $^4$He.\cite{beauvoisPRB16}

The correlation fluctuations are to be determined from the generalization of the
Rayleigh-Ritz variational principle to time-dependent wave functions which states that
minimization of the action integral
\begin{align}
  S = \int \diff{t} \left\langle \Psi(t) \left\vert \hamiltonian - i \hbar \partial_t \right\vert \Psi(t) \right\rangle
  \text{,}
\label{eq:Saction}
\end{align}
is equivalent to solving the many-body Schrödinger equation, see e.\,g.\ Ref.~\onlinecite{KermanKoonin}.
The resulting Euler-Lagrange
equations for $\deltau_\alpha$ and $\deltau_{\alpha \beta}$ are linearized in the spirit of linear
response theory where the perturbation is assumed to be small.  Using the uniform limit approximation
for the pair distribution function and the convolution approximation for the 3-body distribution
function, we can cast the linearized Euler-Lagrange
equations into an equation relating the density response $\Delta\rho_\alpha(\omega)$ (i.\,e.\ change
of the density with respect to the ground state density) to a perturbation
of frequency $\omega$,
\begin{align}
  \Delta\rho_\alpha(\rv_\alpha,\omega)
= \sum_\beta\int\! \diff{\rv}_\beta\, \chi_{\alpha \beta}(\rv_\alpha,\rv_\beta,\omega)\, V_\beta(\rv_\beta,\omega)\,,
\end{align}
which defines the density response matrix $\chi$.  For a homogeneous system, $\chi$ is translationally invariant,
$\chi_{\alpha \beta}(\rv_\alpha,\rv_\beta,\omega)=\chi_{\alpha \beta}(\rv_\alpha-\rv_\beta,\omega)$
such that the integral relation becomes
an algebraic relation in momentum space (see eq.~(\ref{eq:Deltarho}) in the appendix),
\begin{align}
  \Delta\rho_\alpha(\kv,\omega)
= \sum_\beta \chi_{\alpha \beta}(\kv,\omega)\, V_\beta(\kv,\omega)\,.
\end{align}

The density response matrix is related to the absorption spectrum which measures the work $A$
done on the system by the perturbing fields $V_\beta$.
$A$ can be calculated as~\cite{kadanoffAnnPhys63,Forster75}
\begin{align}
  A =
  \int\! {\diff{\kv}\over (2\pi)^2} \int {\omega \diff{\omega}\over 2\pi}\ A(\kv,\omega)\,.
\label{eq:A}
\end{align}
$A(\kv,\omega)$ is the absorption density (i.\,e.\ spectrum) for the momentum
and frequency component $V_\beta(\kv,\omega)$ of the perturbation,
\begin{align}
  A(\kv,\omega) =
  \sum_{\alpha,\beta}
  V^*_\alpha(\kv,\omega)\ \chi''_{\alpha\beta}(\kv,\omega)\ V_\beta(\kv,\omega)
\label{eq:Akw}
\end{align}
with
\begin{align}
  \chi''_{\alpha\beta}(\kv,\omega) = {\chi_{\alpha\beta}^{\phantom{*}}(\kv,\omega)-\chi_{\beta\alpha}^*(\kv,\omega)\over 2i}\,.
\end{align}
Note that if only one layer $x$ is perturbed, i.\,e.\ $V_\beta(\kv,\omega)\sim \delta_{\beta,x}$, then the absorption
spectrum is proportional to just the diagonal element $\chi''_{xx}(\kv,\omega)=\Im\chi_{xx}(\kv,\omega)$.  The density in
layers other than layer $x$ is of course perturbed, too, but these fluctuations do not contribute to
the work done by $V_x$ on layer $x$.

Probing layers individually is an experimental challenge.  It seems more feasible to perturb all layers with
the same external field, $V_\beta(\kv,\omega)=V(\kv,\omega)$, where $\kv$ is parallel to the layers.
If we set the strength of the perturbation to unity, $V(\kv,\omega)=1$,
eq.~(\ref{eq:A}) shows that the absorption spectrum $A_0(\kv,\omega)$ for a given wave number $\kv$ and frequency $\omega$ is
\begin{align}
A_0(\kv,\omega)=\sum_{\alpha,\beta}\chi''_{\alpha\beta}(\kv,\omega)\,.
\label{eq:A0}
\end{align}

In section~\ref{sec:results} below, we consider only symmetric arrangements of $L$ layers, where the system
is invariant under reflection about a mirror plane parallel to the layers (coinciding with the middle layer
if $L$ is uneven).  In this case, the eigenmodes are either symmetric or antisymmetric with respect to
the mirror plane.  If we probe all layers with the same external field,
only the symmetric modes couple to this perturbation, with the
lowest modes coupled most strongly.  Stronger coupling to higher modes as well as coupling to antisymmetric modes is achieved by
angling the wave vector $\kv$ of the perturbation against the layer plane by an angle $\theta$.  This leads to an additional
factor $e^{ikd\beta\sin\theta}$ in $V_\beta(\kv,\omega)$, where $d$ is the layer separation; thus
the perturbation strength becomes layer-dependent.  We denote the correponding absorption spectrum
$A_\theta(\kv,\omega)$.

\section{Results for Dipolar Multilayers}
\label{sec:results}

\subsection{Ground State}
\label{ssec:gs}

Before we can investigate the dynamics, we need to calculate the
ground state properties that enter the calculation of the linear response $\chi''$, namely the
static structure factor matrix $S_{\alpha\beta}(\kv)$, eq.~(\ref{eq:Shom}), and from that
the Bijl-Feynman (BF) states via eq.~(\ref{eq:feynmanhom}).

We consider two cases of multi-layer dipolar Bose condensates: the first case is an
ideal limiting case, where the total
density $\rho_{\rm tot}$ is split evenly between the $L$ layers, leading to partial
densities $\rho_\alpha=\rho_{\rm tot}/L$.  The second case is closer to a practical
experimental implementation, where the outermost layers have lower densities than
more central layers.  We assume a symmetric parabolic distribution of partial densities.
In both cases, we chose a total density
of $\rho_{\rm tot}r_D^2=1$ which for a {\em single} layer gives rise to strong correlations,
-- too strong to allow a mean field description~\cite{dipolePRL09}--, but not strong
enough for roton excitations.  For all following results we choose a system
of $L=8$ coupled layers.  Hence for equal partial densities, we have $\rho_\alpha r_D^2=1/8$
for all $\alpha$.  For the parabolic distribution of partial denstities we
have $\rho_\alpha r_D^2 = \rho_{8-\alpha+1} r_D^2 = 0.0436; 0.1134; 0.1599; 0.1831$ for $\alpha=1;2;3;4$.

\begin{figure}[ht]
\hfill\hfill\includegraphics*[width=0.15\textwidth]{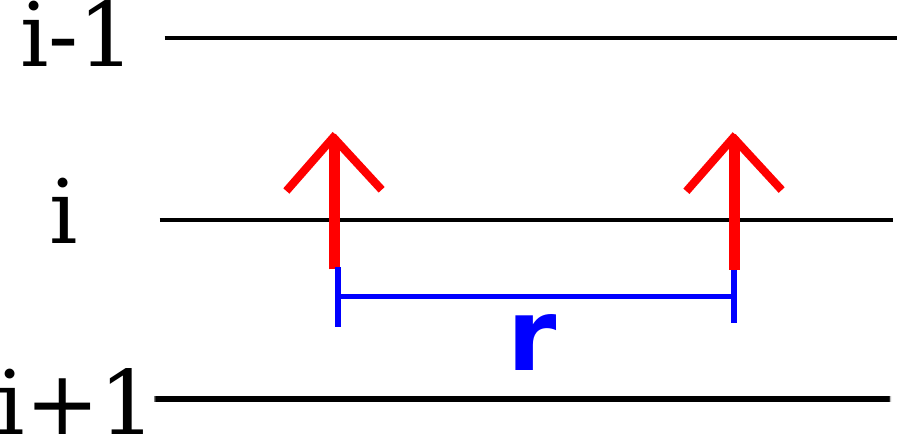}\hfill\includegraphics*[width=0.15\textwidth]{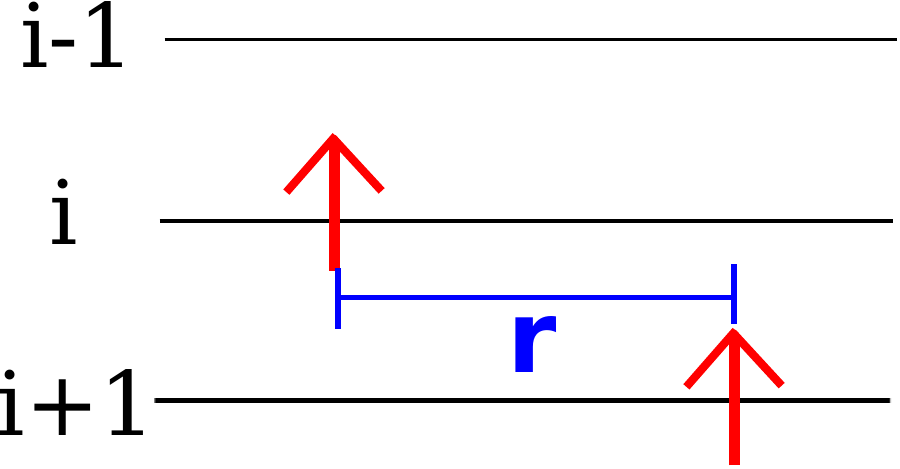}\hfill\smallskip

\includegraphics*[width=0.45\textwidth]{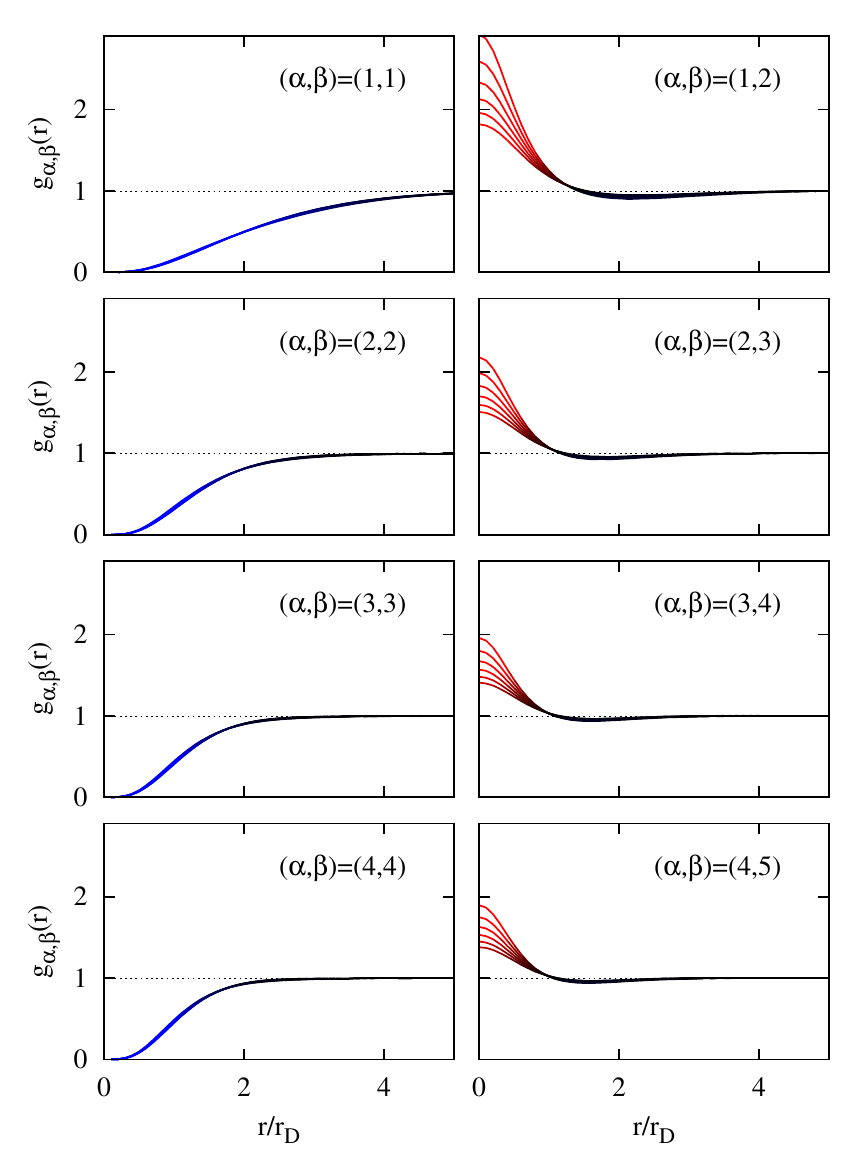}
\caption{
  $g_{\alpha\beta}(r)$ matrix for $L=8$ layers with parabolic density distribution for $L=8$ layers.
  $(\alpha,\beta)$ is indicated in each panel, where the left and right column show correlations in
  the same layer and between adjacent layers, respectively, illustrated by the two sketches above the two
  columns.  The distance between layers
  is decreased, $d/r_D=1.000;0.950;0.902;0.857;0.815;$ and $0.774$. This leads to the increasing
  peak height in the correlations between adjacent layers, while the intralayer correlations have
  a negligible dependence on $d$.
}
\label{FIG:gpar}
\end{figure}

Fig.~\ref{FIG:gpar} shows elements $g_{\alpha\beta}(r)$ of the pair distribution 
matrix for a parabolic distribution of
partial densities, respectively.  The left columns shows $g_{\alpha\beta}(r)$ for intralayer
correlations, $\alpha=\beta$, and the right columns shows $g_{\alpha\beta}(r)$ for correlations
between adjacent layers, $\alpha=\beta-1$.  Several curves are shown in each case, corresponding
to decreasing values for the distances $d$ between layers ($d/r_D=1.000;0.950;0.902;0.857;0.815;0.774$).
If we decrease $d$ further, the variational Jastrow-Feenberg ansatz (\ref{eq:Psi0}) does not lead to a
stable ground state: the numerical optimization according to the HNC-EL equations does not converge
to a meaningful result. The reason is that dipoles of adjacent layers will form bound states which
is not accounted for in the ansatz (\ref{eq:Psi0}).  For bilayers,
the pairing effect has been well studied by quantum Monte Carlo simulations~\cite{maciaPRA14,filinovPRA16}
and is not the subject of this work.

The intralayer correlations $g_{\alpha\alpha}(r)$ in the left column are dominated by the correlations hole due to
the repulsive dipole-dipole interaction. They are essentially independent of the distance $d$ (all curves
coincide).  Conversely, the correlations between different layers are dominated by a peak at $r=0$ due to
the attractive dipole-dipole interaction for a head-to-tail arrangement.  The interlayer correlation
peaks between particles in adjacent layers increase with decreasing $d$.\footnote{this is also true between
layers separated by several layers (not shown); $g_{\alpha\beta}(r)$ $|\alpha-\beta|=2,3,\dots$ looks qualitatively
the same as for adjacent layers, but overall the peaks are lower}
Note that while $d$ is decreased only slightly in Fig.~\ref{FIG:gpar},
the peak at $r=0$ of $g_{\alpha\beta}(r)$ for adjacent layers, $\alpha=\beta-1$, increases by more than a factor of two.
Furthermore the higher peak for outer layers demonstrates that interlayer correlations due to dipole coupling
become stronger for {\em lower} partial densities.\cite{hebenstreitPRA16}

For a uniform distribution of partial densities, $g_{\alpha\beta}(r)$ looks qualitatively similar to
Fig.~\ref{FIG:gpar} (correlation hole for intralayer correlations, correlation peak for interlayer
correlations), there is no dependence on $\alpha$ for $g_{\alpha\alpha}(r)$ since all layers have
the same density.  For $g_{\alpha\beta}(r)$ with $\alpha=\beta-1$ there is only a very small dependence on
the location of the adjacent layer.

\subsection{Dynamics}

The multilayered dipolar Bose gas shall now be weakly perturbed to probe the excitations.
As discussed above, we consider the same perturbation with wave vector $\kv$ and frequency $\omega$
for all layers.  For the moment we assume that $\kv$ is parallel to the layers, $\theta=0$.
We use the ground state results for the 8-layer system with either a uniform or a parabolic
distribution of partial densities as described above.  We choose a
small layer distance $d/r_D=0.774$ where the interlayer correlations for both distributions are
strong, but the ground state is still stable against formation of bound states between dipoles
in different layers.  We also compare with the dynamics of uncoupled layers, i.\,e.\ of
layers with a large $d$.  Excitations that have infinite lifetime (within the CBF
approximation) appear as $\delta$-peaks in $A_0(\kv,\omega)$. Especially for lower wave
numbers, these peaks with zero linewidth carry the main spectral strength and correspond
to excitations where the density oscillations move in phase in all layers.
In order to visualize the spectral weight of peaks with zero linewidth in $A_0(\kv,\omega)$, we introduce
as usual a small artificial linewidth of $0.015 E_D$.  In addition to the sharp peaks,
$A_0(\kv,\omega)$ exhibits broader regions that can be attributed to simultaneous
excitation of two excitations.

\begin{figure}[ht]
\includegraphics*[width=0.9\columnwidth]{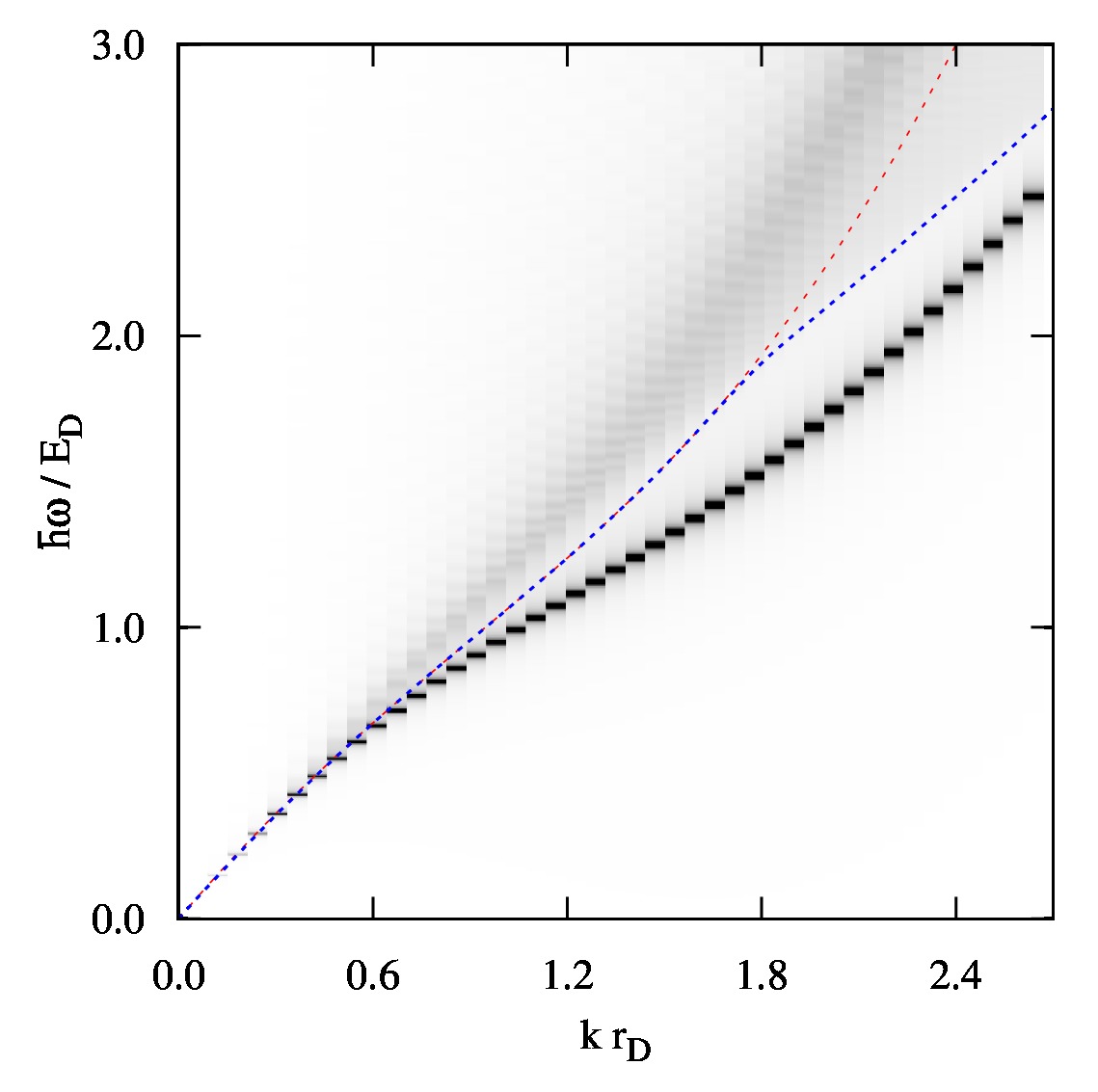}
\caption{
Absorption spectrum $A_0(\kv,\omega)$ for 8 layers of Bose dipoles, each having
the same partial density $\rho_\alpha r_D^2=1/8$.  The layers are uncoupled.
Due to the identity of all layers, $A_0(\kv,\omega)$ has only a single
excitation branch.  The red dotted line is the dispersion in BF approximation
and the dashed blue line is the decay threshold $b_{11}(k)$
discussed in the text.
}
\label{FIG:L8d10}
%
\includegraphics*[width=0.9\columnwidth]{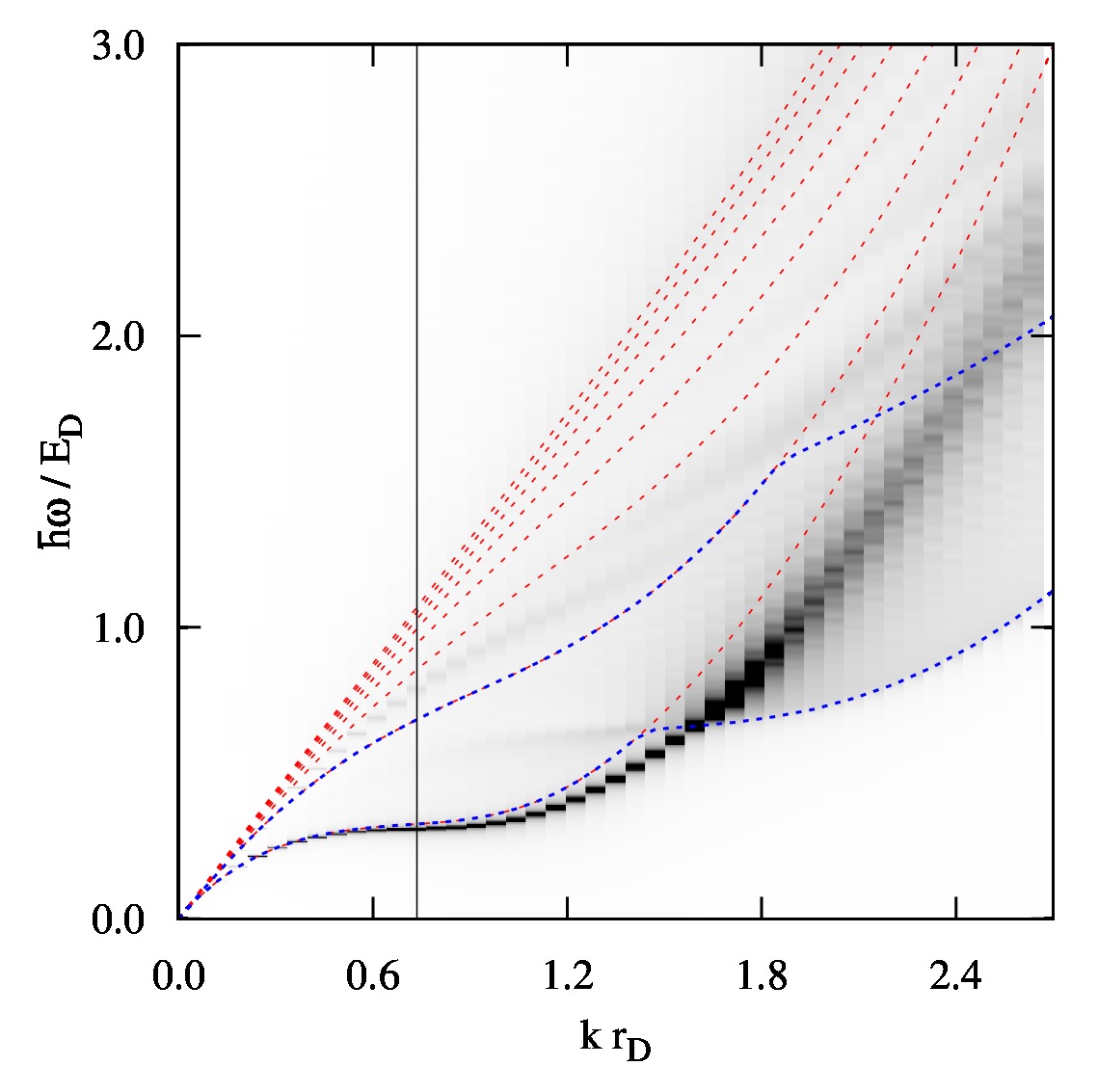}
\caption{
Same as Fig.~\ref{FIG:L8d10} for strong interlayer coupling, achieved for a small layer separation $d/r_D=0.774$.
The lower and upper dashed blue lines are the decay thresholds $b_{11}(k)$ and $b_{22}(k)$
discussed in the text.
}
\label{FIG:L8}
\end{figure}

The absorption spectrum $A_0(\kv,\omega)$ for $L=8$ layers with {\em equal} partial densities
$\rho_\alpha=\rho_{\rm tot}/L$ is shown in Figs.~\ref{FIG:L8d10} and \ref{FIG:L8}.  Fig.~\ref{FIG:L8d10} shows
$A_0(\kv,\omega)$ for $d/r_D=10$, i.\,e.\ effectively for 8 independent layers without
coupling between them.  Since all
layers are identical, $A_0(\kv,\omega)$ exhibits only a single, undamped dispersion,
with a small, strongly damped multi-excitation peak above the dominant peak of the
excitation.  The red dotted line indicates the single $\delta$-peak following from the \BFA .

The more interesing case of strongly coupled layers separated
by $d/r_D=0.774$ is shown in Fig.~\ref{FIG:L8}.
Already in \BFA\ the excitation energies are split due to the interlayer DDI.
The absorption spectrum $A_0(\kv,\omega)$, however, is dominated by a main
peak at much lower energies than in the uncoupled case.  Hence, despite the splitting
of the excitation energies due to the interlayer coupling, the system still
approximately behaves like a single 2D layer, at least for perturbations with
wave vector parallel to the layers.  The dispersion, defined by following the
main peak in $A_0(\kv,\omega)$, has zero slope around
$kr_D\approx 0.8$, hence the system is on the verge of ``rotonization''.
We discuss roton, i.\,e.\ local minima in the dispersion, further below.
The vertical line in Fig.~\ref{FIG:L8} indicates the cut of $A_0(\kv,\omega)$
for $k r_D=0.736$ shown further below.

There are weak signals from higher
excitations, but they have very small spectral weight and would probably be hard to
detect in experiments.  For example, a weak, damped peak can be discerned above the
main peak, increasing in strength until it merges with the main peak around $kr_D\approx 1.6$.
This can be attributed to a double-excitation which has a non-negligibe cross section if
the density of states is high -- as it is the case for excitations in the range $kr_D\approx 0.6\dots 0.9$
where the slope of the dispersion is very small.  The high density
of states for these excitation leads to an increased susceptibilty for exciting
two modes simultaneously.

Following the main peak in Fig.~\ref{FIG:L8} beyond the almost-roton to larger $k$, we 
observe that damping sets in at about $kr_D\approx 1.6$.  Below that threshold, the
peak has zero linewidth (within the CBF approximation), the broadening seen in
Fig.~\ref{FIG:L8} is the artificial broadening necessary to vizualize not just
the dispersion relation, but also the spectral weight, as mentioned above.
Above $kr_D\approx 1.6$, the excitation energy is above the threshold where
decay into two excitations of lower energy becomes kinematically allowed.
As can be seen from the energy denominator in the expression for the self energy in the BF basis,
$\bar\Sigma_{mn}(\kv_m,\omega)$ (eq.~(\ref{eq:sigmahom})), in the CBF approximation
an excitation of energy $\hbar\omega$ decays into two BF modes.  The threshold
for decay into BF modes $n$ and $m$ is given by
$$
b_{nm}(k)={\rm min}_q[\varepsilon_{n}(q)+\varepsilon_{m}(|\kv-\qq|)]\;.
$$
It is intuitively clear that the decay mechanism should not be the creation of two BF modes,
but rather should create
modes with an energy determined selfconsistently, i.\,e.\ including the self energy
correction.  Since the self energy lowers the excitation energies, the actual
decay threshold will be slightly lower than predicted by the CBF approximation.
As mentioned earlier, this deficiency of CBF could be cured by including triplet
correlations, as has been derived for homogeneous single-component Bose
systems.~\cite{campbellPRB15,beauvoisPRB16}  Whether the inclusion of triplet correlations
in multi-component systems is feasible, is being investigated.

\begin{figure}[ht]
\includegraphics*[width=0.9\columnwidth]{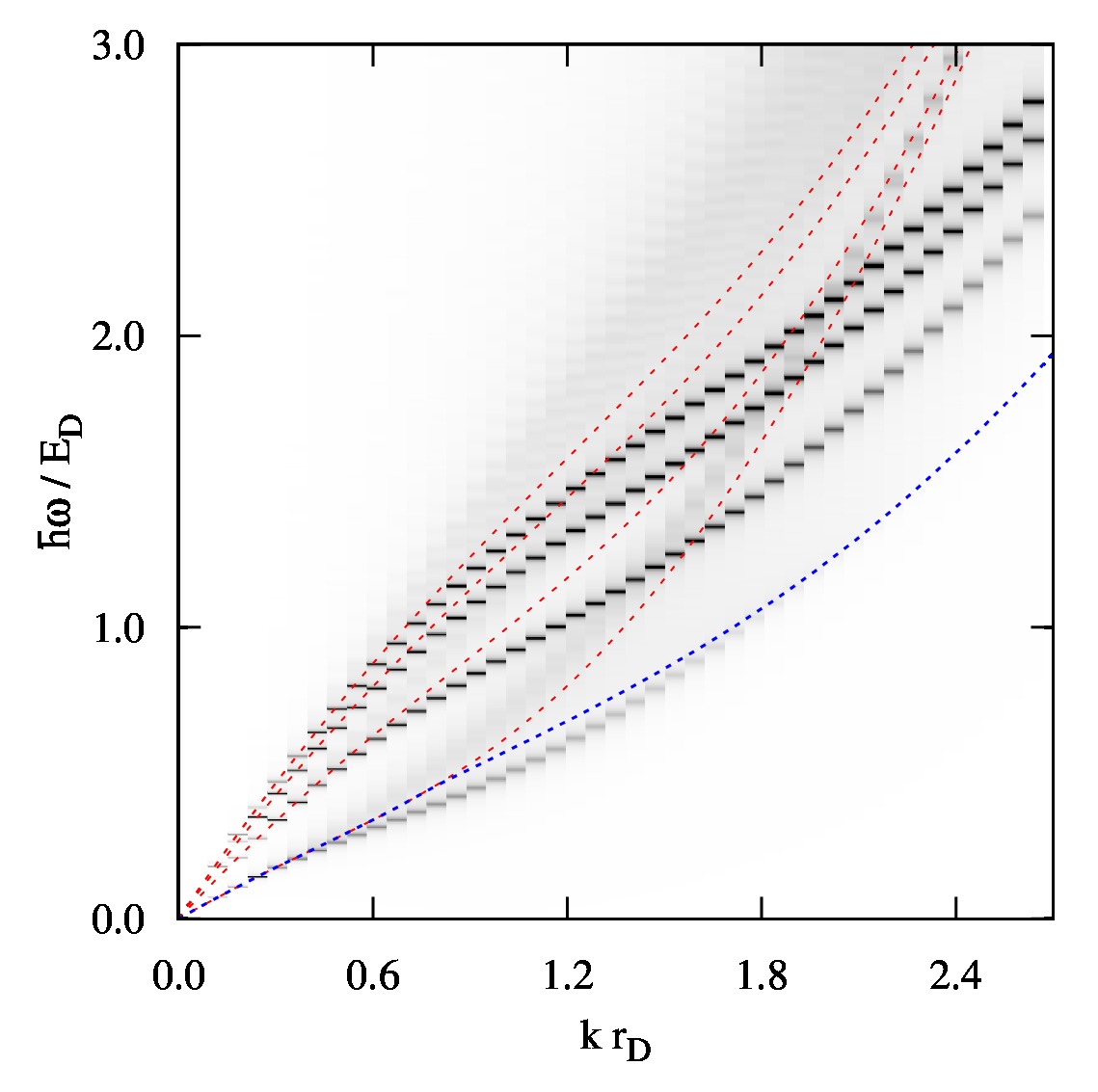}
\caption{
Absorption spectrum $A_0(\kv,\omega)$ for 8 layers of Bose dipoles, with a parabolic
profile for the partial densities $\rho_\alpha$. The layers are uncoupled.  Due to
symmetry, there are 4 independent excitations visible in $A_0(\kv,\omega)$, where modes of
higher energy and spectral weights are excitations of layers with higher partial density.
}
\label{FIG:L8pard10}
%
\includegraphics*[width=0.9\columnwidth]{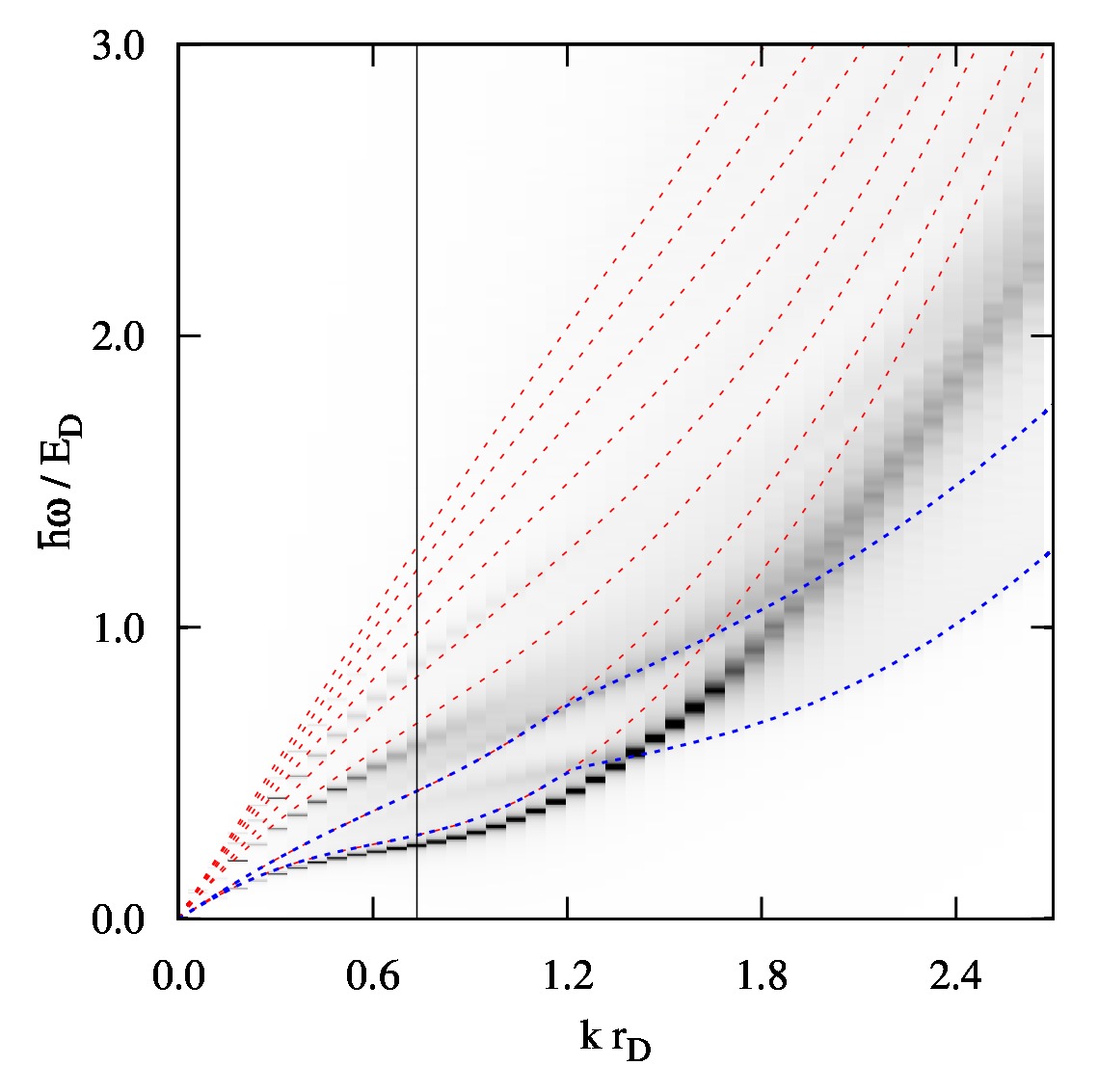}
\caption{
Same as Fig.~\ref{FIG:L8pard10} for strong coupling, achieved for a small layer separation $d/r_D=0.774$.
The lower and upper dashed blue lines are the decay thresholds $b_{11}(k)$ and $b_{22}(k)$
discussed in the text.
Most of the spectral weight collapses into a single collective mode, involving all layers.
}
\label{FIG:L8par}
\end{figure}

The decay threshold into the lowest BF states, $b_{11}(k)$ is shown in Fig.~\ref{FIG:L8}
as the lower dashed blue line.  The higher dashed blue line is $b_{22}(k)$, the
decay threshold into the two next BF state.  Crossing $b_{22}(k)$ leads to additional
damping.  Note that $b_{12}(k)=b_{21}(k)$ does not play a role here: perturbating all
layers equally, $V_\beta(\kv,\omega)=V(\kv,\omega)$, is of course symmetric with
respect to reflection about the mirror plane, see discussion of symmetry above, and
can thus excite only symmetric excitations.
The BF states are alternatingly symmetric and antisymmetric, with the lowest
being symmetry.  The selection rule, implicit in the self energy, is such that
a symmetric excitation can decay only into two symmetric or two antisymmetric
BF modes, but not into two BF modes of mixed symmetries.

The absorption spectrum $A_0(\kv,\omega)$ for $L=8$ layers for the more realistic
case of {\em non-equal} partial densities $\rho_\alpha$, following a parabolic distribution,
is shown in Figs.~\ref{FIG:L8pard10} and \ref{FIG:L8par}.  Again, we show first
the spectrum $A_0(\kv,\omega)$ for a layer distance of $d/r_D=10$, i.\,e.\ for uncoupled layers,
in Fig.~\ref{FIG:L8pard10}.
Since we have an ensemble of 4 different layers now (two for each of the 4 different partial
densities) $A_0(\kv,\omega)$ exhibits the dispersions of 4 different excitations,
the lower ones corresponding to lower partial densities.  $A_0(\kv,\omega)$ is simply the
sum of these 4 absorption spectra, in the absence of interlayer coupling.
Lifetime broadening of an exciation in layer $\alpha$
is possible only via decay into two excitations in the same layer $\alpha$.  The lowest
excitation becomes strongly damped above $kr_D\approx 2$, while for the other excitations, damping sets in for
higher $k$, outside the range shown in Fig.~\ref{FIG:L8pard10}.  Higher excitation
have higher spectral weight, which is simply because of the higher density, as seen in
eq.~(\ref{eq:chihom}).

Fig.~\ref{FIG:L8par} shows $A_0(\kv,\omega)$ for strongly coupled layers separated
by $d/r_D=0.774$.  Unlike in the case of equal partial densities, where the dominant
effect of the coupling was to lower the main
peak in $A_0(\kv,\omega)$, the coupling has much stronger effect in the case of non-equal
partial densities.  Without coupling we have $L/2=4$ peaks in $A_0(\kv,\omega)$,
(Fig.~\ref{FIG:L8pard10}),
with the most spectral weight at higher energy (coming from excitation in higher density layers).
The coupling collapses most of the spectral weight into a {\em single} peak:
the $L=8$ layers act like a single effective layer.  As for layers with equal density $\rho_\alpha$,
the dispersion relation of the main peak is pulled to lower energy by the coupling.
Higher energy features of $A_0(\kv,\omega)$ above the main peak are now a bit
more pronounced than
in Fig.~\ref{FIG:L8}.  The absorption due to exciting higher even modes
(again, we stress that odd modes cannot be excited by a symmetric perturbation)
is barely visible for small wave numbers, $kr_D<1$, but they are completely damped for
higher $k$.  For $kr_D>1$, the two-excitation dispersion is visible, see explanation above.
As in Fig.~\ref{FIG:L8} the two dashed blue lines are the two
decay thresholds $b_{11}(k)$ and $b_{22}(k)$.
Again, we observe that the main excitation peak becomes damped when it crosses $b_{11}(k)$
(decay into two symmetrix modes), and is damped further when it crosses $b_{22}(k)$.
(decay into two antisymmetric modes).

\begin{figure}[ht]
\includegraphics*[width=0.9\columnwidth]{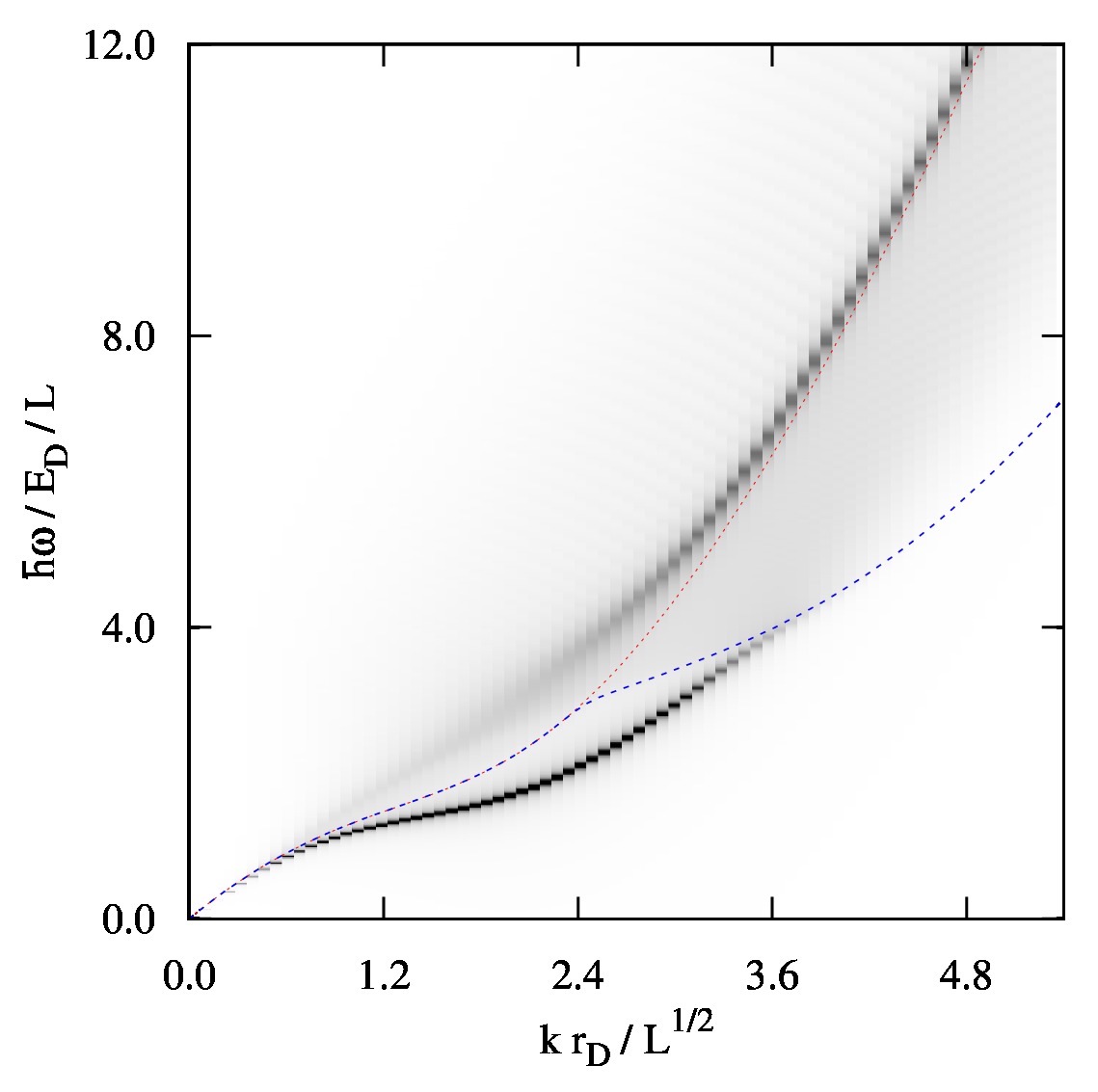}
\caption{
Absorption spectra of a single dipole layer with a density $\rho r_D^2=1$, corresponding
to the total density $\rho_{\rm tot}=\sum_\alpha\rho_\alpha$ of the 8-layer systems.
}
\label{FIG:L1}
\end{figure}

The collapse of most of the spectral weight in the absorption spectrum $A_0(\kv,\omega)$
into a single collective mode begs the question whether there is an equivalent
{\em single} two-dimensional layer system leading to very similar spectra as those shown in
Figs.~\ref{FIG:L8} and \ref{FIG:L8par} for the strongly coupled 8-layer systems.
We calculate $A_0(\kv,\omega)$ of a single dipole layer with a density given
by the total density of the 8-layer systems,
$\rho \equiv \rho_{\rm tot}=\sum_\alpha\rho_\alpha$, where $\rho_{\rm tot}r_D^2=1$.
Single layers of dipolar Bose gases have been studied extensively in the
past~\cite{buechlerPRL07,astraPRL07,filinovPRL10,maciaPRA11,maciaPRL12,filinovPRA12},
and the CBF spectra for various densities have been presented in Ref.~\onlinecite{dipolePRL09}.
In Fig.~\ref{FIG:L1} we show the absorption spectrum $A_0(\kv,\omega)$, which
for a single layer is synonymous with the dynamic structure function $S(\kv,\omega)$.
Putting all dipoles into a single layer increases the density by a factor of 8
with respect to the single layer density, at least in the case where all partial
densities are equal, $\rho_\alpha r_D^2=1/8$.  In order to account for the higher
density, we rescaled the energies and momenta by $L^{-1}$ and $L^{-1/2}$, respectively.

The shape of the undamped part of the dispersion (up to $kr_D\approx 3.6$)
is indeed similar to the dispersion in Figs.~\ref{FIG:L8} and \ref{FIG:L8par}.
However, even with the rescaled momenta and energies, the absorption spectrum
of a single layer shown in Fig.~\ref{FIG:L1} is not comparable with the
strongly coupled multilayer spectra in Figs.~\ref{FIG:L8} and \ref{FIG:L8par}.
The excitation energies associated with the sharp main peaks is about a factor of 4 larger
and the momenta about a factor of 2.  More importantly, the dispersion relation defined
by the sharp main peaks in Fig.~\ref{FIG:L1} actually loses spectral weight
as it approaches the damping threshold (dashed blue line). The spectral weight is
gradually shifted to a damped excitation at higher energy slightly above
the BF dispersion relation.  When the undamped lower branch of the dispersion
reaches the damping threshold, it completely vanishes.
Hence we conclude that the absorption spectrum
of strongly coupled multilayers, despite being dominated by a
single peak carrying most of the spectral weight, is not just a (rescaled)
single layer spectrum, with all partial densities packed into a single layer.

An interesting aspect of dipolar Bose gases in a layer geometry is the conjecture
to obtain a dispersion with roton excitations, i.\,e.\ a well-defined local minimum
at a certain wave number $k_r$.\cite{odellPRL03,santosPRL03}  The appearance of
rotons may be favorable if the DDI attraction for head-to-tail arrangment
of two dipoles becomes sufficiently large.  Using CBF, single layers
and bilayers of finite width have been studied, and indeed the dispersion of the
excitations can become ``flat'' (zero slope) for some $k_r$, and even exhibits
a very shallow roton minimum.\cite{hufnaglPRL11,hufnaglPRA13}  However, no deep roton
was found in these CBF-based studies.  Before a well-developed, deep roton
could form, the variational ansatz for the ground state was always numerically
unstable.  Coming back to the present case of 2D multilayers, we observe the same
trend.  Here we increase the coupling strength by decreasing the distance $d$ between layers;
the collective excitations become softer and the slope of the dispersion relation
vanishes (Fig.~\ref{FIG:L8}) or becomes very small (Fig.~\ref{FIG:L8par}) around $k_rr_D\approx 0.7$.
However, as discussed in section~\ref{ssec:gs},
the Jastrow-Feenberg ansatz (\ref{eq:Psi0}) for the ground state cannot be
numerically optimized anymore if $d$ is decreased further, since it does
not account for the formation of bound states of dipoles in different layers.
Hence, just like in the previous CBF calculations of layers of finite
width,\cite{hufnaglPRL11,hufnaglPRA13} we do not find a well-developed, deep roton
-- at least if the CBF spectra are based on a variational ground
state without interlayer bound states.

We caution that CBF is an approximation and the corrections
of 3-body and higher correlations may well be important in the regime
of strong interlayer coupling, despite the low partial densities considered
in this work.  But we note that the absence of rotons with a well-defined minimum
is consistent with recent results for the excitations
of dipolar bilayers obtained with quantum Monte Carlo
simulations.\cite{filinovPRA16,astraPRA16}  It was found that rotons appeared when dipoles
form bound states (dimers), or if the density in each layer is so high that
rotons are present due to the intralayer dipole-dipole repulsion, regardless
of the interlayer coupling.

\begin{figure}[ht]
\includegraphics*[width=0.45\textwidth]{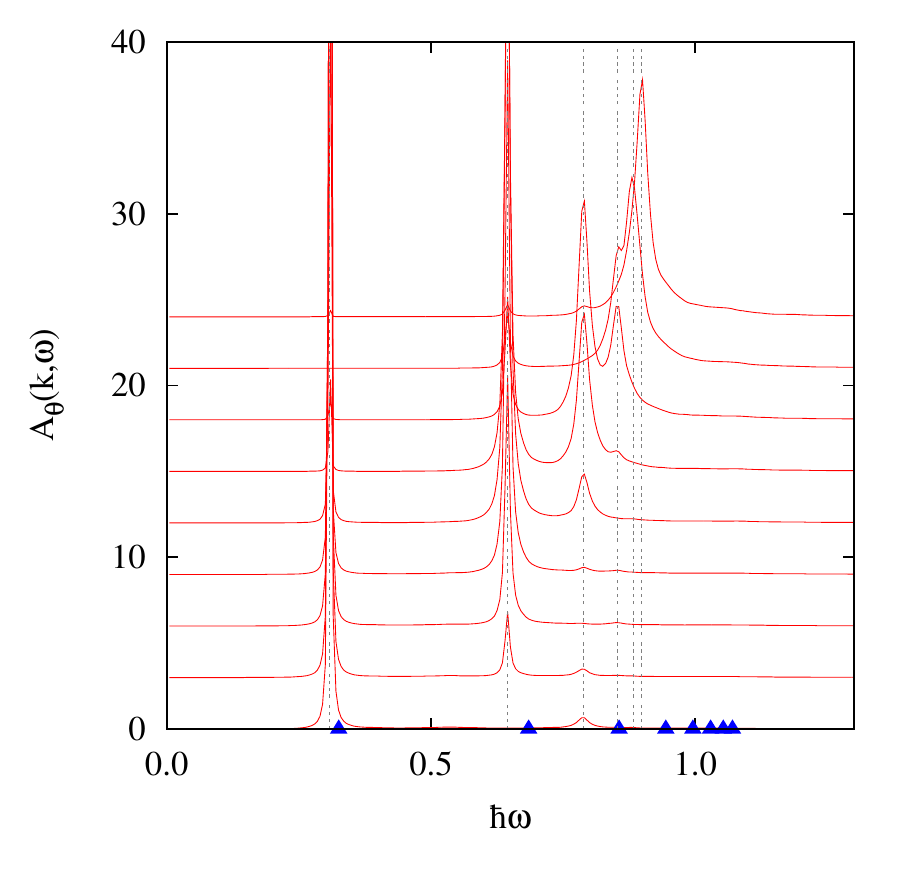}
\caption{
Absorption spectrum $A_\theta(\kv,\omega)$ for 8 layers of Bose dipoles, each having
the same partial density $\rho_\alpha r_D^2=1/8$, for a layer separation $d/r_D=0.774$.
The angle $\theta$ between layer plane
and the wave vector $\kv$ of the perturbation is increased in steps of $10^\circ$ between
$0^\circ$ and $80^\circ$, where the curves are offset by an amound proportional
to $\theta$.
The projection $k=|\kv|\cos\theta$ of $\kv$ on the plane was chosen
$k r_D=0.736$.
}
\label{FIG:L8angle}
\end{figure}

\begin{figure}[ht]
\includegraphics*[width=0.45\textwidth]{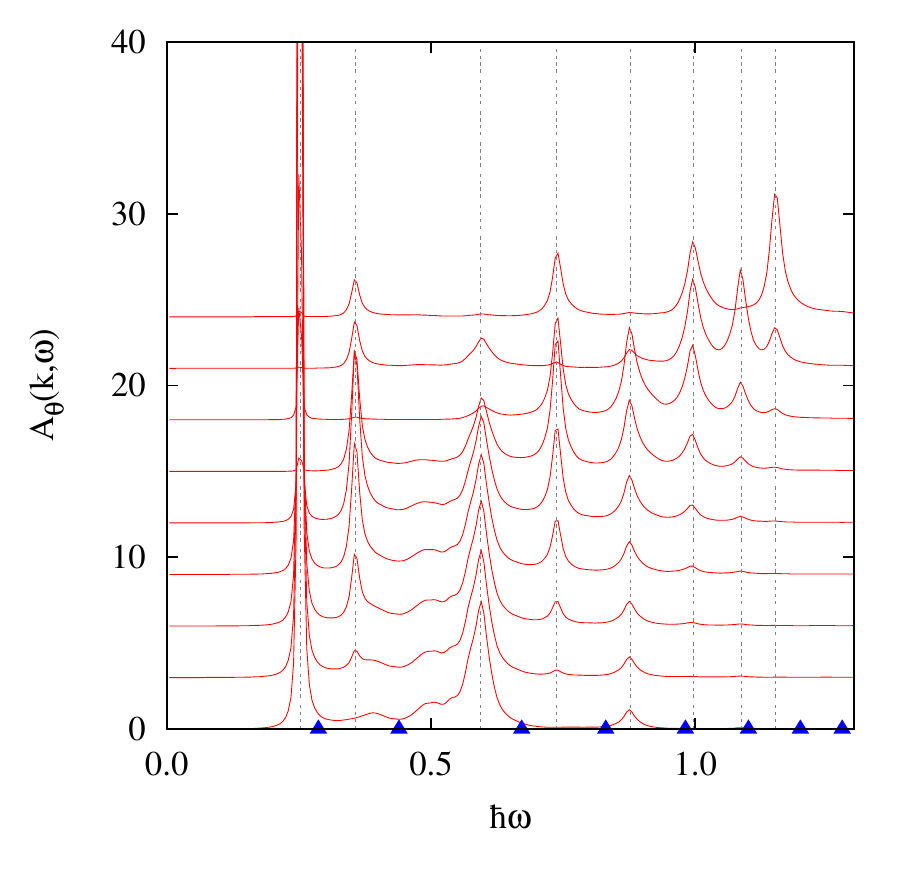}
\caption{
Same as Fig.~\ref{FIG:L8angle}, but for the parabolic density profile.
}
\label{FIG:L8parangle}
\end{figure}

The absorption spectra presented so far show that, due to the interlayer
coupling, the dynamic response of a multilayered Bose gas of dipoles
is dominated by a single main peak carrying most of the spectral weight
-- provided that we restrict the
perturbation to wave vectors parallel to the layers.
Only the lowest mode has appreciable spectral weight if the perturbation is
parallel to the layers.  But higher states can indeed be
probed if we perturb the multilayered dipolar Bose gas with wave vectors
that have a finite angle with respect to the layers plane.  For bilayers
with finite thickness, this was shown in Ref.~\onlinecite{hufnaglPRA13}.
As we have argued at the end of section~\ref{sec:theory}, a non-parallel wave vector
causes a perturbation with a different phase in each layer, while the
amplitude is still the same in all layers.

In Fig.~\ref{FIG:L8angle} (equal partial densities)
and Fig.~\ref{FIG:L8parangle} (parabolic partial densities) we show the absorption
spectrum $A_\theta(\kv,\omega)$ for $k r_D=0.736$, where $k=|\kv|\cos\theta$ now denotes the
projection of $\kv$ on the layer plane (only the parallel component
of $\kv$ is a good quantum number) and $\theta$ is the angle
of $\kv$ with respect to the layer plane.
A parallel component of $k r_D=0.736$ was chosen because it lies in the interesting
range where the slope of the lowest mode becomes small.
The lowest curves are $A_\theta(\kv,\omega)$ for an angle $\theta=0^\circ$, i.\,e.\
reproducing the results from above for the wave number $k r_D=0.736$
(the curves for $\theta=0^\circ$ correspond to the cuts indicated by
vertical lines in Figs.~\ref{FIG:L8} and \ref{FIG:L8par}).
$\theta$ is then increased in steps of 10$^\circ$ up to $\theta=80^\circ$.
The spectra for these angles are offset by an amount proportional to $\theta$
for better visibility.
As $\theta$ is increased, higher modes indeed gain spectral weight, including
antisymmetric modes that are forbidden transitions for perturbations
with even symmetry, as is the case for $\theta=0^\circ$.
Single excitation peaks are marked by dashed vertical lines in
Figs.~\ref{FIG:L8angle} and \ref{FIG:L8parangle}; additional broad peaks
with small spectral weight in Fig.~\ref{FIG:L8parangle} are the multi-excitation peaks discussed above.
The triangles on the abscissa mark the corresponding excitation energies
obtained in the \BFA\ for comparison.  For this choice of $k$,
all modes but the lowest have a finite
linewidth (the lowest has only the artificial linewidth used for visualization),
since they can decay into two modes.  For example the second mode (which is
the first antisymmetric mode) can decay into an antisymmetric mode and the first (symmetric) mode.
Note that for equal partial densities higher states
are very closely spaced, as can be seen from the excitation obtained
in the \BFA , which neglects damping effects. Therefore the broadening
of the lines due to damping in the full CBF spectra makes it impossible
to discern the three highest states in Fig.~\ref{FIG:L8angle}.

\section{Conclusions and Outlook}

We generalized the CBF method to multicomponent many-body systems, where
all components are bosons.  In our derivation we use the linear
response approach to CBF, which leads to coupled linear equations relating
perturbations of a Bose species $\alpha$ and the response of the
density of species $\beta$.  The density response matrix $\chi_{\alpha\beta}$
relating perturbation and response can be expressed in terms of Bijl-Feynman
excitations and a self energy matrix which accounts for the renormalization
of excitation energies as well as dissipation due to the coupling of
Bijl-Feynman excitations.

As an application, we consider a dipolar Bose gas
in a one-dimensional optical trap deep enough to suppress
tunneling and to render each layer effectively two-dimensional.  Due to the long range
of the DDI, the layers are not just an ensemble of independent
2D Bose gases, but are coupled via the DDI acting between layers.
We considered two cases of 8 dipolar layers: equal partial densities in each
layer, and a parabolic distribution of partial densities.  The dynamic
response of these two multilayered dipolar gases is quite different in
the absence of interlayer coupling. However, when we include the interlayer
coupling they both have qualitatively similar absorption spectra, with a single
excitation carrying most of the spectral weight, provided we apply a
uniform perturbation to all layers.  If on the other hand we tilt the
wave vector of the perturbation such that the perturbation has different relative phases
in different layers, the absorption spectra reveals also information about
higher lying states.
Finally we note that, although the dispersion relation of the lowest
mode exhibits a flat part, we did not find evidence of a well-defined,
deep roton minimum.

In our work, we did not allow for pairing between particles of different
species.  In the case of coupled dipolar multilayers, pairing of dipoles
in different layers can occur when the dipole length $r_D$ is sufficiently
large.  For a bilayer system, Monte Carlo simulations have predicted
a phase transition from a single particle superfluid to a ``pair superfluid''.\cite{maciaPRA14}
Regarding the dynamics in a paired state, estimates of the
excitation energies based on imaginary time correlations with Monte Carlo simulations
show that the antisymmetric excitation (so-called spin mode) is gapped in the
pair-superfluid state.\cite{filinovPRA16,astraPRA16}
Therefore, it would be interesting to generalize
the CBF method further to allow such a pairing.  Reconstructions
of the energy spectra from imaginary time data obtained by Monte Carlo simulation
usually have a low resolution due to the inherent ill-posed nature of the problem.
Although it is not exact, CBF provides high resolution response functions,
showing details such as multi-excitations peaks, and predicting the linewidth
of peaks.  Multicomponent CBF would provide valuable additional
insight into the dynamics of pair superfluid bilayers that cannot be afforded
by exact Monte Carlo simulations.  We note that CBF accepts as input ground state
quantities obtained with any available method, e.\,g.\ obtained with QMC.

For single-component homogeneous Bose systems, recent improvements of the self energy
have been achieved by including time-dependent triplet correlations.
\cite{campbellPRB15,beauvoisPRB16}
We therefore plan to include triplet correlations at least in some approximate way
such that the computational effort is still feasible.

\begin{acknowledgments}
We acknowledge financial support by the Austrian Science Fund FWF (grant No.\ 23535),
and discussions with Eckhard Krotscheck, Ferran Mazzanti, Jordi Boronat, and Gregory Astrakharchik.
\end{acknowledgments}


\begin{appendix}

\newlength{\dgrmBaseLength}
\setlength{\dgrmBaseLength}{8mm}
\newlength{\dgrmBaseHeight}
\setlength{\dgrmBaseHeight}{\dgrmBaseLength * \real{0.866025404}} 
\newlength{\dgrmBaseLengthQuarter}
\setlength{\dgrmBaseLengthQuarter}{\dgrmBaseLength / 4}
\newlength{\dgrmBaseLengthHalf}
\setlength{\dgrmBaseLengthHalf}{\dgrmBaseLength / 2}
\newlength{\dgrmBaseLengthThreeQuarter}
\setlength{\dgrmBaseLengthThreeQuarter}{\dgrmBaseLength * 3 / 4}
\newlength{\dgrmCircleRadius}
\setlength{\dgrmCircleRadius}{\dgrmBaseLength / 12}
\newlength{\dgrmBaseHeightHalf}
\setlength{\dgrmBaseHeightHalf}{\dgrmBaseHeight / 2}
\newlength{\dgrmBaseHeightThird}
\setlength{\dgrmBaseHeightThird}{\dgrmBaseHeight / 3}
\newlength{\dgrmBaseHeightSixth}
\setlength{\dgrmBaseHeightSixth}{\dgrmBaseHeight / 6}
\newlength{\dgrmBaseHeightTwoThird}
\setlength{\dgrmBaseHeightTwoThird}{\dgrmBaseHeight * 2 / 3}

\newcommand{\dgrmwww}{
    \vcenter{
        \hbox{
            \begin{tikzpicture}
                \tikzstyle{every node}=[font=\tiny]
                \coordinate (A) at (0,0);
		        \coordinate (B) at (\dgrmBaseLength,0);
		        \coordinate (C) at (\dgrmBaseLengthHalf,\dgrmBaseHeight);
		        \fill[white] (A) circle (\dgrmCircleRadius);
		        \fill[white] (B) circle (\dgrmCircleRadius);
		        \fill[white] (C) circle (\dgrmCircleRadius);
		        \draw (A) circle (\dgrmCircleRadius);
		        \draw (B) circle (\dgrmCircleRadius);
		        \draw (C) circle (\dgrmCircleRadius);
            \end{tikzpicture}
        }
    }
}
\newcommand{\dgrmwwwl}{
    \vcenter{
        \hbox{
            \begin{tikzpicture}
                \tikzstyle{every node}=[font=\tiny]
                \coordinate (A) at (0,0);
		        \coordinate (B) at (\dgrmBaseLength,0);
		        \coordinate (C) at (\dgrmBaseLengthHalf,\dgrmBaseHeight);
		        \draw[style=solid] (A) -- (B);
		        \fill[white] (A) circle (\dgrmCircleRadius);
		        \fill[white] (B) circle (\dgrmCircleRadius);
		        \fill[white] (C) circle (\dgrmCircleRadius);
		        \draw (A) circle (\dgrmCircleRadius);
		        \draw (B) circle (\dgrmCircleRadius);
		        \draw (C) circle (\dgrmCircleRadius);
            \end{tikzpicture}
        }
    }
}
\newcommand{\dgrmwwwll}{
    \vcenter{
        \hbox{
            \begin{tikzpicture}
                \tikzstyle{every node}=[font=\tiny]
                \coordinate (A) at (0,0);
		        \coordinate (B) at (\dgrmBaseLength,0);
		        \coordinate (C) at (\dgrmBaseLengthHalf,\dgrmBaseHeight);
		        \draw[style=solid] (A) -- (C);
		        \draw[style=solid] (B) -- (C);
		        \fill[white] (A) circle (\dgrmCircleRadius);
		        \fill[white] (B) circle (\dgrmCircleRadius);
		        \fill[white] (C) circle (\dgrmCircleRadius);
		        \draw (A) circle (\dgrmCircleRadius);
		        \draw (B) circle (\dgrmCircleRadius);
		        \draw (C) circle (\dgrmCircleRadius);
            \end{tikzpicture}
        }
    }
}
\newcommand{\dgrmwwwblll}{
    \vcenter{
        \hbox{
            \begin{tikzpicture}
                \tikzstyle{every node}=[font=\tiny]
                \coordinate (A) at (0,0);
		        \coordinate (B) at (\dgrmBaseLength,0);
		        \coordinate (C) at (\dgrmBaseLengthHalf,\dgrmBaseHeight);
		        \coordinate (M) at (\dgrmBaseLengthHalf,\dgrmBaseHeightThird);
		        \draw[style=solid] (A) -- (M);
		        \draw[style=solid] (B) -- (M);
		        \draw[style=solid] (C) -- (M);
		        \fill[white] (A) circle (\dgrmCircleRadius);
		        \fill[white] (B) circle (\dgrmCircleRadius);
		        \fill[white] (C) circle (\dgrmCircleRadius);
		        \fill (M) circle (\dgrmCircleRadius);
		        \draw (A) circle (\dgrmCircleRadius);
		        \draw (B) circle (\dgrmCircleRadius);
		        \draw (C) circle (\dgrmCircleRadius);
            \end{tikzpicture}
        }
    }
}
\newcommand{\dgrmwwwt}{
    \vcenter{
        \hbox{
            \begin{tikzpicture}
                \tikzstyle{every node}=[font=\tiny]
                \coordinate (A) at (0,0);
		        \coordinate (B) at (\dgrmBaseLength,0);
		        \coordinate (C) at (\dgrmBaseLengthHalf,\dgrmBaseHeight);
		        \draw[fill=gray] (A) -- (B) -- (C) -- (A);
		        \fill[white] (A) circle (\dgrmCircleRadius);
		        \fill[white] (B) circle (\dgrmCircleRadius);
		        \fill[white] (C) circle (\dgrmCircleRadius);
		        \draw (A) circle (\dgrmCircleRadius);
		        \draw (B) circle (\dgrmCircleRadius);
		        \draw (C) circle (\dgrmCircleRadius);
            \end{tikzpicture}
        }
    }
}
\newcommand{\dgrmwwwblt}{
    \vcenter{
        \hbox{
            \begin{tikzpicture}
                \tikzstyle{every node}=[font=\tiny]
                \coordinate (A) at (0,0);
		        \coordinate (B) at (\dgrmBaseLength,0);
		        \coordinate (C) at (\dgrmBaseLengthHalf,\dgrmBaseHeight);
		        \coordinate (A2) at (\dgrmBaseLengthHalf,\dgrmBaseHeightThird);
		        \draw[style=solid] (A2) -- (C);
		        \draw[fill=gray] (A) -- (B) -- (A2) -- (A);
		        \fill (A2) circle (\dgrmCircleRadius);
		        \fill[white] (A) circle (\dgrmCircleRadius);
		        \fill[white] (B) circle (\dgrmCircleRadius);
		        \fill[white] (C) circle (\dgrmCircleRadius);
		        \draw (A) circle (\dgrmCircleRadius);
		        \draw (B) circle (\dgrmCircleRadius);
		        \draw (C) circle (\dgrmCircleRadius);
            \end{tikzpicture}
        }
    }
}
\newcommand{\dgrmwwwbbllt}{
    \vcenter{
        \hbox{
            \begin{tikzpicture}
                \tikzstyle{every node}=[font=\tiny]
                \coordinate (A) at (0,0);
		        \coordinate (B) at (\dgrmBaseLength,0);
		        \coordinate (C) at (\dgrmBaseLengthHalf,\dgrmBaseHeight);
		        \coordinate (A2) at (\dgrmBaseLengthQuarter,\dgrmBaseHeightHalf);
		        \coordinate (B2) at (\dgrmBaseLengthThreeQuarter,\dgrmBaseHeightHalf);
		        \draw[style=solid] (A) -- (A2);
		        \draw[style=solid] (B) -- (B2);
		        \draw[fill=gray] (A2) -- (B2) -- (C) -- (A2);
		        \fill (A2) circle (\dgrmCircleRadius);
		        \fill (B2) circle (\dgrmCircleRadius);
		        \fill[white] (A) circle (\dgrmCircleRadius);
		        \fill[white] (B) circle (\dgrmCircleRadius);
		        \fill[white] (C) circle (\dgrmCircleRadius);
		        \draw (A) circle (\dgrmCircleRadius);
		        \draw (B) circle (\dgrmCircleRadius);
		        \draw (C) circle (\dgrmCircleRadius);
            \end{tikzpicture}
        }
    }
}
\newcommand{\dgrmwwwbbblllt}{
    \vcenter{
        \hbox{
            \begin{tikzpicture}
                \tikzstyle{every node}=[font=\tiny]
                \coordinate (A) at (0,0);
		        \coordinate (B) at (\dgrmBaseLength,0);
		        \coordinate (C) at (\dgrmBaseLengthHalf,\dgrmBaseHeight);
		        \coordinate (A2) at (\dgrmBaseLengthQuarter,\dgrmBaseHeightSixth);
		        \coordinate (B2) at (\dgrmBaseLengthThreeQuarter,\dgrmBaseHeightSixth);
		        \coordinate (C2) at (\dgrmBaseLengthHalf,\dgrmBaseHeightTwoThird);
		        \draw[style=solid] (A) -- (A2);
		        \draw[style=solid] (B) -- (B2);
		        \draw[style=solid] (C) -- (C2);
		        \draw[fill=gray] (A2) -- (B2) -- (C2) -- (A2);
		        \fill (A2) circle (\dgrmCircleRadius);
		        \fill (B2) circle (\dgrmCircleRadius);
		        \fill (C2) circle (\dgrmCircleRadius);
		        \fill[white] (A) circle (\dgrmCircleRadius);
		        \fill[white] (B) circle (\dgrmCircleRadius);
		        \fill[white] (C) circle (\dgrmCircleRadius);
		        \draw (A) circle (\dgrmCircleRadius);
		        \draw (B) circle (\dgrmCircleRadius);
		        \draw (C) circle (\dgrmCircleRadius);
            \end{tikzpicture}
        }
    }
}

\newcommand{\dgrmwwwlabels}[3]{
    \vcenter{
        \hbox{
            \begin{tikzpicture}
                \tikzstyle{every node}=[font=\tiny]
                \coordinate[label={[label distance=0mm]left:$#1$}] (A) at (0,0);
		        \coordinate[label={[label distance=0mm]right:$#2$}] (B) at (\dgrmBaseLength,0);
		        \coordinate[label={[label distance=0mm]left:$#3$}] (C) at (\dgrmBaseLengthHalf,\dgrmBaseHeight);
		        \fill[white] (A) circle (\dgrmCircleRadius);
		        \fill[white] (B) circle (\dgrmCircleRadius);
		        \fill[white] (C) circle (\dgrmCircleRadius);
		        \draw (A) circle (\dgrmCircleRadius);
		        \draw (B) circle (\dgrmCircleRadius);
		        \draw (C) circle (\dgrmCircleRadius);
            \end{tikzpicture}
        }
    }
}
\newcommand{\dgrmwwwllabels}[3]{
    \vcenter{
        \hbox{
            \begin{tikzpicture}
                \tikzstyle{every node}=[font=\tiny]
                \coordinate[label={[label distance=0mm]left:$#1$}] (A) at (0,0);
		        \coordinate[label={[label distance=0mm]right:$#2$}] (B) at (\dgrmBaseLength,0);
		        \coordinate[label={[label distance=0mm]left:$#3$}] (C) at (\dgrmBaseLengthHalf,\dgrmBaseHeight);
		        \draw[style=solid] (A) -- (B);
		        \fill[white] (A) circle (\dgrmCircleRadius);
		        \fill[white] (B) circle (\dgrmCircleRadius);
		        \fill[white] (C) circle (\dgrmCircleRadius);
		        \draw (A) circle (\dgrmCircleRadius);
		        \draw (B) circle (\dgrmCircleRadius);
		        \draw (C) circle (\dgrmCircleRadius);
            \end{tikzpicture}
        }
    }
}
\newcommand{\dgrmwwwlllabels}[3]{
    \vcenter{
        \hbox{
            \begin{tikzpicture}
                \tikzstyle{every node}=[font=\tiny]
                \coordinate[label={[label distance=0mm]left:$#1$}] (A) at (0,0);
		        \coordinate[label={[label distance=0mm]right:$#2$}] (B) at (\dgrmBaseLength,0);
		        \coordinate[label={[label distance=0mm]left:$#3$}] (C) at (\dgrmBaseLengthHalf,\dgrmBaseHeight);
		        \draw[style=solid] (A) -- (C);
		        \draw[style=solid] (B) -- (C);
		        \fill[white] (A) circle (\dgrmCircleRadius);
		        \fill[white] (B) circle (\dgrmCircleRadius);
		        \fill[white] (C) circle (\dgrmCircleRadius);
		        \draw (A) circle (\dgrmCircleRadius);
		        \draw (B) circle (\dgrmCircleRadius);
		        \draw (C) circle (\dgrmCircleRadius);
            \end{tikzpicture}
        }
    }
}
\newcommand{\dgrmwwwbllllabels}[3]{
    \vcenter{
        \hbox{
            \begin{tikzpicture}
                \tikzstyle{every node}=[font=\tiny]
                \coordinate[label={[label distance=0mm]left:$#1$}] (A) at (0,0);
		        \coordinate[label={[label distance=0mm]right:$#2$}] (B) at (\dgrmBaseLength,0);
		        \coordinate[label={[label distance=0mm]left:$#3$}] (C) at (\dgrmBaseLengthHalf,\dgrmBaseHeight);
		        \coordinate (M) at (\dgrmBaseLengthHalf,\dgrmBaseHeightThird);
		        \draw[style=solid] (A) -- (M);
		        \draw[style=solid] (B) -- (M);
		        \draw[style=solid] (C) -- (M);
		        \fill[white] (A) circle (\dgrmCircleRadius);
		        \fill[white] (B) circle (\dgrmCircleRadius);
		        \fill[white] (C) circle (\dgrmCircleRadius);
		        \fill (M) circle (\dgrmCircleRadius);
		        \draw (A) circle (\dgrmCircleRadius);
		        \draw (B) circle (\dgrmCircleRadius);
		        \draw (C) circle (\dgrmCircleRadius);
            \end{tikzpicture}
        }
    }
}
\newcommand{\dgrmwwwtlabels}[3]{
    \vcenter{
        \hbox{
            \begin{tikzpicture}
                \tikzstyle{every node}=[font=\tiny]
                \coordinate[label={[label distance=0mm]left:$#1$}] (A) at (0,0);
		        \coordinate[label={[label distance=0mm]right:$#2$}] (B) at (\dgrmBaseLength,0);
		        \coordinate[label={[label distance=0mm]left:$#3$}] (C) at (\dgrmBaseLengthHalf,\dgrmBaseHeight);
		        \draw[fill=gray] (A) -- (B) -- (C) -- (A);
		        \fill[white] (A) circle (\dgrmCircleRadius);
		        \fill[white] (B) circle (\dgrmCircleRadius);
		        \fill[white] (C) circle (\dgrmCircleRadius);
		        \draw (A) circle (\dgrmCircleRadius);
		        \draw (B) circle (\dgrmCircleRadius);
		        \draw (C) circle (\dgrmCircleRadius);
            \end{tikzpicture}
        }
    }
}
\newcommand{\dgrmwwwbltlabels}[3]{
    \vcenter{
        \hbox{
            \begin{tikzpicture}
                \tikzstyle{every node}=[font=\tiny]
                \coordinate[label={[label distance=0mm]left:$#1$}] (A) at (0,0);
		        \coordinate[label={[label distance=0mm]right:$#2$}] (B) at (\dgrmBaseLength,0);
		        \coordinate[label={[label distance=0mm]left:$#3$}] (C) at (\dgrmBaseLengthHalf,\dgrmBaseHeight);
		        \coordinate (A2) at (\dgrmBaseLengthHalf,\dgrmBaseHeightThird);
		        \draw[style=solid] (A2) -- (C);
		        \draw[fill=gray] (A) -- (B) -- (A2) -- (A);
		        \fill (A2) circle (\dgrmCircleRadius);
		        \fill[white] (A) circle (\dgrmCircleRadius);
		        \fill[white] (B) circle (\dgrmCircleRadius);
		        \fill[white] (C) circle (\dgrmCircleRadius);
		        \draw (A) circle (\dgrmCircleRadius);
		        \draw (B) circle (\dgrmCircleRadius);
		        \draw (C) circle (\dgrmCircleRadius);
            \end{tikzpicture}
        }
    }
}
\newcommand{\dgrmwwwbblltlabels}[3]{
    \vcenter{
        \hbox{
            \begin{tikzpicture}
                \tikzstyle{every node}=[font=\tiny]
                \coordinate[label={[label distance=0mm]left:$#1$}] (A) at (0,0);
		        \coordinate[label={[label distance=0mm]right:$#2$}] (B) at (\dgrmBaseLength,0);
		        \coordinate[label={[label distance=0mm]left:$#3$}] (C) at (\dgrmBaseLengthHalf,\dgrmBaseHeight);
		        \coordinate (A2) at (\dgrmBaseLengthQuarter,\dgrmBaseHeightHalf);
		        \coordinate (B2) at (\dgrmBaseLengthThreeQuarter,\dgrmBaseHeightHalf);
		        \draw[style=solid] (A) -- (A2);
		        \draw[style=solid] (B) -- (B2);
		        \draw[fill=gray] (A2) -- (B2) -- (C) -- (A2);
		        \fill (A2) circle (\dgrmCircleRadius);
		        \fill (B2) circle (\dgrmCircleRadius);
		        \fill[white] (A) circle (\dgrmCircleRadius);
		        \fill[white] (B) circle (\dgrmCircleRadius);
		        \fill[white] (C) circle (\dgrmCircleRadius);
		        \draw (A) circle (\dgrmCircleRadius);
		        \draw (B) circle (\dgrmCircleRadius);
		        \draw (C) circle (\dgrmCircleRadius);
            \end{tikzpicture}
        }
    }
}
\newcommand{\dgrmwwwbbbllltlabels}[3]{
    \vcenter{
        \hbox{
            \begin{tikzpicture}
                \tikzstyle{every node}=[font=\tiny]
                \coordinate[label={[label distance=0mm]left:$#1$}] (A) at (0,0);
		        \coordinate[label={[label distance=0mm]right:$#2$}] (B) at (\dgrmBaseLength,0);
		        \coordinate[label={[label distance=0mm]left:$#3$}] (C) at (\dgrmBaseLengthHalf,\dgrmBaseHeight);
		        \coordinate (A2) at (\dgrmBaseLengthQuarter,\dgrmBaseHeightSixth);
  		        \coordinate (B2) at (\dgrmBaseLengthThreeQuarter,\dgrmBaseHeightSixth);
  		        \coordinate (C2) at (\dgrmBaseLengthHalf,\dgrmBaseHeightTwoThird);
		        \draw[style=solid] (A) -- (A2);
		        \draw[style=solid] (B) -- (B2);
		        \draw[style=solid] (C) -- (C2);
		        \draw[fill=gray] (A2) -- (B2) -- (C2) -- (A2);
		        \fill (A2) circle (\dgrmCircleRadius);
		        \fill (B2) circle (\dgrmCircleRadius);
		        \fill (C2) circle (\dgrmCircleRadius);
		        \fill[white] (A) circle (\dgrmCircleRadius);
		        \fill[white] (B) circle (\dgrmCircleRadius);
		        \fill[white] (C) circle (\dgrmCircleRadius);
		        \draw (A) circle (\dgrmCircleRadius);
		        \draw (B) circle (\dgrmCircleRadius);
		        \draw (C) circle (\dgrmCircleRadius);
            \end{tikzpicture}
        }
    }
}

\newcommand{\labelleddgrmplus}{\!\!\!\!\!+\!\!\!\!\!}

\section{Derivation of Multicomponent CBF}

Here we derive the CBF approximation for the linear response.
We use the CBF formulation of Ref.~\onlinecite{clementsPRB96} for single component Bose systems
and generalize it to multicomponent Bose systems.

In some of the definitions that follow we integrate over all
particle coordinates except for one or two particles, and since all particles
of the same component are identical, we can choose e.\,g.\ the first particle
without loss of generality.  Therefore
we introduce the abbreviations $\diff{\tau_\alpha}$ and $\diff{\tau_{\alpha \beta}}$.
They are given by the product of differentials $\prod_{\alpha, j}\diff{r_{\alpha,j}}$
where $\diff{r_{\alpha,1}}$ is omitted and
$\diff{r_{\alpha,1}} \diff{r_{\beta,1+\delta_{\alpha \beta}}}$ is omitted,
respectively.  Furthermore, we abbreviate the combinatorial factors
\begin{gather}
  \Pi_\alpha \equiv \frac{N_\alpha!}{\left(N_\alpha - 1\right)!}
  \quad \text{and} \quad
  \Pi_{\alpha \beta} \equiv 
  \begin{cases}
    \frac{N_\alpha !}{(N_\alpha - 2)!} & \text{if } \alpha = \beta \\
    \Pi_\alpha \Pi_\beta & \text{if } \alpha \neq \beta
  \end{cases}
  \text{.}
\end{gather}
From now on we use component index vectors $\vec{\alpha}$, e.\,g. $\vec{\alpha} = (\alpha)$ or $\vec{\alpha} = (\alpha, \beta)$ and omit the dependence on coordinates since for each component $\alpha$ there is always a coordinate $\rv_{\alpha}$.  For example, we abbreviate $\deltau_{\alpha \beta}(\rv_{\alpha}, \rv_{\beta}, t) \equiv \deltau_{\alpha \beta}(t)$.

We define the complex-valued density fluctuations
\begin{align}
  \deltarho_{\vec{\alpha}} \equiv \Pi_{\vec{\alpha}} \int \diff{\tau_{\vec{\alpha}}} 
    |\psi_0|^2 \left( \excitationOp - \left\langle \excitationOp \right\rangle_{\! 0} \right)
\end{align}
where $\delta U$ is defined in eq.~(\ref{eq:deltaU}) and
$\left\langle \excitationOp \right\rangle_{0} \equiv \left\langle \psi_0 \left\vert \excitationOp \right\vert \psi_0 \right\rangle$ is the ground state expectation value.  We also define the complex valued current densities
\begin{align}
  \vec{j}_{\vec{\alpha}}
  \equiv \frac{\hbar \Pi_{\vec{\alpha}}}{2 i m_{\alpha_1}} \int \diff{\tau_{\vec{\alpha}}} |\psi_0|^2 \nabla_{\alpha_1} \excitationOp
  \text{.}
\end{align}
Note that $\Re \delta \rho_{\vec{\alpha}}$ ($\Re \vec{j}_{\vec{\alpha}}$) are the physical density fluctuations (current densities)
to linear order in the perturbation. Furthermore we define
\begin{align}
  D_{\vec{\alpha}}
  \equiv \frac{2 \Pi_{\vec{\alpha}}}{i \hbar} \int \diff{\tau_{\vec{\alpha}}} |\psi_0|^2 \left(
    \deltaHamiltonian - \left\langle \deltaHamiltonian \right\rangle_{\!0}
  \right)
\end{align}
and minimize the action $S$, eq.~(\ref{eq:Saction}), with respect to $\delta u_{\alpha}$ and $\delta u_{\alpha \beta}$, which,
to linear order in the perturbation, gives the following coupled equations of motion (EOMs):
\begin{gather}
  \label{eq:eom1}
  \nabla_\alpha \cdot \vec{j}_\alpha + \deltarhoDot_\alpha  = D_{\alpha} + \order{\delta U^2} 
  \\
  \label{eq:eom2}
  \nabla_\alpha \cdot \vec{j}_{\alpha \beta} + \nabla_\beta \cdot \vec{j}_{\beta \alpha} + \deltarhoDot_{\alpha \beta}
  = D_{\alpha \beta} + \order{\delta U^2}
\end{gather}
The ground state $n$-body densities are given by
\begin{align}
  \rho_{\vec{\alpha}} \equiv \Pi_{\vec{\alpha}} \int \diff{\tau_{\vec{\alpha}}} |\psi_0(\Rv)|^2
  \text{,}
\end{align}
the corresponding two- and three-body distribution functions $g_{\alpha \beta}$ and $g_{\alpha \beta \gamma}$,
as well as $h_{\alpha \beta}$ and $h_{\alpha \beta \gamma}$, are
\begin{gather}
  \rho_{\alpha \beta} \equiv g_{\alpha \beta} \rho_\alpha \rho_\beta \equiv \left( 1 + h_{\alpha \beta} \right) \rho_\alpha \rho_\beta \\
  \rho_{\alpha \beta \gamma} \equiv g_{\alpha \beta \gamma} \rho_\alpha \rho_\beta \rho_\gamma
    \equiv (1+h_{\alpha \beta \gamma}) \rho_\alpha \rho_\beta \rho_\gamma
\end{gather}
We introduce the tilde-notation $\tilde{f}_{\vec{\alpha}} \equiv \sqrt{\rho_{\alpha_1} \rho_{\alpha_2} \ldots} f_{\vec{\alpha}}$,
but with the exceptions
$\deltarhoTilde_\alpha = \frac{\deltarho_\alpha}{\sqrt{\rho_\alpha}}$,
$\vec{\tilde{j}}_\alpha = \frac{\vec{j}_\alpha}{\sqrt{\rho_\alpha}}$,
$\widetilde{D}_\alpha = \frac{D_\alpha}{\sqrt{\rho_\alpha}}$
and $\widetilde{D}_{\alpha \beta} = \frac{D_{\alpha \beta}}{\sqrt{\rho_\alpha \rho_\beta}}$
and use the shorthand notation
\begin{align}
  \label{eq:sumintnotation}
  \sum\limits_{\alpha} \int\! \diff{\rv_\alpha} \longrightarrow  \sumint_{\alpha}
  \text{.}
\end{align}
The static structure factor of the ground state can be written as
\begin{align}
  S_{\alpha \beta} = \delta_{\alpha \beta} \delta\left( \rv_\alpha - \rv_\beta \right) + \tilde{h}_{\alpha \beta}
  \text{.}
\end{align}
With all these definitions, we can write down the terms in the EOMs (\ref{eq:eom1}) and (\ref{eq:eom2})
\begin{widetext}
\begin{align}
  \widetilde{D}_\alpha
  &= \frac{2}{i \hbar} \sumint_{\beta} S_{\alpha \beta} \potentialTilde_\beta \\
  \widetilde{D}_{\alpha \beta}
  &= \frac{2}{i \hbar} \left(
    \tilde{g}_{\alpha \beta} \left( \potential_\alpha + \potential_\beta \right)
    + \sumint_{\gamma} \left( \tilde{g}_{\alpha \beta \gamma} - \sqrt{\rho_\gamma} \tilde{g}_{\alpha \beta} \right) \potentialTilde_\gamma
  \right)
  \label{eq:Dalphabeta}
  \\
  \vec{j}_{\alpha}
  &= \frac{\hbar \rho_\alpha}{2 i m_\alpha} \left(
    \nabla_\alpha \deltau_\alpha
    + \sumint_{\beta} \rho_\beta g_{\alpha \beta} \nabla_\alpha \deltau_{\alpha \beta}
  \right)
  \\
  \vec{\tilde{j}}_{\alpha \beta}
  &= \frac{\hbar}{2 i m_\alpha} \left(
    \tilde{g}_{\alpha \beta} \nabla_\alpha \left( \deltau_\alpha + \deltau_{\alpha \beta} \right)
    + \sumint_{\gamma} \sqrt{\rho_\gamma} \tilde{g}_{\alpha \beta \gamma} \nabla_\alpha \deltau_{\alpha \gamma}
  \right)
  \\
  \deltarhoTilde_\alpha
  &= \sumint_{\beta} \left(
    S_{\alpha \beta} \deltauTilde_\beta
    + \sqrt{\rho_\beta} g_{\alpha \beta} \deltauTilde_{\alpha \beta}
    + \frac{1}{2} \sumint_{\gamma} \left( \tilde{g}_{\alpha \beta \gamma} - \sqrt{\rho_\alpha} \tilde{g}_{\beta \gamma} \right)
      \deltauTilde_{\beta \gamma}
  \right)
\end{align}
\end{widetext}
The two-body density fluctuations are expanded to linear order in the fluctuations,
\begin{align}
  \label{eq:deltarhoDotTwo}
  \delta \dot{\rho}_{\alpha \beta}
  = g_{\alpha \beta} \left( \rho_\alpha \deltarhoDot_\beta + \rho_\beta \deltarhoDot_\alpha \right)
    + \rho_\alpha \rho_\beta \dot{g}_{\alpha \beta}
    + \order{\delta U^2}
\end{align}
The three-body distribution function is approximated using the convolution approximation\cite{FeenbergBook}.
Using the notation of the Meyer cluster diagrams~\cite{Hansen},
\begin{align}
  \notag
  g_{\alpha \beta \gamma}
  &\approx
  \dgrmwww + 
  \dgrmwwwl +
  \dgrmwwwll +
  \dgrmwwwblll \\
  &+ \dgrmwwwt +
  \dgrmwwwblt +
  \dgrmwwwbbllt +
  \dgrmwwwbbblllt
  \text{,}
\end{align}
where a white dot stands for an external variable $\alpha$
and a black dot stands for an internal variable $\alpha$, for which we multiply
with a factor $\rho_\alpha$ and sum over $\alpha$ and integrate over $\qr_\alpha$.
A line connecting $\alpha$ and $\beta$ denotes a factor $h_{\alpha \beta}$
and a shaded triangle connecting $\alpha$, $\beta$ and $\gamma$ stands for the
direct triplet correlation function $X_{\alpha \beta \gamma}$.  They are a straightforward
generalization of the single-component case. \cite{Clements93}
The external variables are not labeled as we sum over all distinct
permutations of $\alpha$, $\beta$ and $\gamma$.  An example would be
\begin{align}
  \dgrmwwwblt = \sumint_{\eta} \rho_\eta \left( 
    h_{\alpha \eta} X_{\eta \beta \gamma} 
    + h_{\beta \eta} X_{\eta \alpha \gamma} 
    + h_{\gamma \eta} X_{\eta \alpha \beta} 
  \right)
  \text{.}
\end{align}
We define the non-nodal part of $g_{\alpha \beta \gamma}$ as
\begin{align}
  Y_{\alpha \beta \gamma}
  \equiv h_{\alpha \beta} h_{\alpha \gamma}
  + \sumint_{\eta,\theta} \frac{S_{\beta \eta} S_{\gamma \theta}}{\sqrt{\rho_\alpha \rho_\beta \rho_\gamma}} \widetilde{X}_{\alpha \eta \theta}
  \text{.}
\end{align}
Applying the convolution approximation in the expression (\ref{eq:Dalphabeta}) for $D_{\alpha \beta}$ and inserting $D_{\alpha}$ into it gives
\begin{align}
  \widetilde{D}_{\alpha \beta}
  &\approx g_{\alpha \beta} \left( \sqrt{\rho_\alpha} \widetilde{D}_\beta + \sqrt{\rho_\beta} \widetilde{D}_\alpha \right)
  + \sumint_{\gamma} \widetilde{Y}_{\gamma \alpha \beta} \widetilde{D}_\gamma
  \text{.}
\end{align}
We eliminate $\deltau_\alpha$ in favor of $\deltarho_\alpha$ in $\vec{j}_\alpha$.
Therefore we rewrite $\deltarhoTilde_\alpha$ as a convolution with $S_{\alpha \beta}$.
\begin{widetext}
\begin{align}
  \deltarhoTilde_\alpha
  &\approx \sumint_{\beta} S_{\alpha \beta} \left(
    \deltauTilde_\beta
    + \sumint_{\gamma} \sqrt{\rho_\gamma} g_{\beta \gamma} \deltauTilde_{\beta \gamma}
    + \frac{1}{2} \sumint_{\gamma,\eta} \widetilde{Y}_{\beta \gamma \eta} \deltauTilde_{\gamma \eta}
  \right)
\end{align}
This relation can be inverted using $S^{-1}_{\alpha \beta} = \delta_{\alpha \beta} \delta(\rv_\alpha - \rv_\beta) - \widetilde{X}_{\alpha \beta}$,
which follows from the Ornstein-Zernike equation (OZE) and where $X_{\alpha \beta}$ is the direct correlation function.
We can reformulate $\vec{j}_{\alpha}$,
\begin{align}
  \vec{j}_\alpha
  &= \frac{\hbar \rho_\alpha}{2 i m_\alpha} \sumint_{\beta} \left(
    \nabla_\alpha \frac{1}{\sqrt{\rho_\alpha}} S^{-1}_{\alpha \beta} \deltarhoTilde_\beta
    - \rho_\beta \deltau_{\alpha \beta} \nabla_\alpha g_{\alpha \beta}
    - \frac{1}{2} \sumint_{\gamma} \rho_\beta \rho_\gamma \deltau_{\beta \gamma} \nabla_\alpha Y_{\alpha \beta \gamma}
  \right)\,.
\end{align}
The next approximation we introduce is the uniform limit approximation (ULA)\cite{FeenbergBook} for terms with $\deltau_{\alpha \beta}$.
\begin{align}
  \deltau_{\alpha \beta} &\approx \deltaX_{\alpha \beta} \\
  g_{\alpha \beta} \nabla_\alpha \deltau_{\alpha \beta} &\approx \nabla_\alpha \deltau_{\alpha \beta} \\
  \left( g_{\alpha \beta \gamma} - g_{\alpha \beta} g_{\alpha \gamma} \right) \nabla_\alpha \deltau_{\alpha \beta}
  &\approx h_{\beta \gamma} \nabla_\alpha \deltau_{\alpha \gamma}
\end{align}
We define $J_{\alpha \beta}$ and apply the ULA,
\begin{align}
  J_{\alpha \beta}
  &\equiv \nabla_\alpha \cdot \left( \vec{j}_{\alpha \beta} - g_{\alpha \beta} \rho_\beta \vec{j}_\alpha \right)
  \ \approx \frac{\hbar \sqrt{\rho_\beta}}{2 i m_\beta} \nabla_\alpha \cdot \rho_\alpha \nabla_\alpha \frac{1}{\sqrt{\rho_\alpha}} \sumint_{\gamma} S_{\beta \gamma} \deltaXTilde_{\alpha \gamma}\,.
\end{align}
Next we reformulate the second EOM \eqref{eq:eom2} by subtracting \eqref{eq:eom1} twice (multiplied with $g_{\alpha \beta}$ and $\rho_\alpha$/$\rho_\beta$) and inserting $\deltarhoDot_{\alpha \beta}$ from \eqref{eq:deltarhoDotTwo}.
\begin{align}
  \label{eq:eom2alternate}
  J_{\alpha \beta} + J_{\beta \alpha}
  + \left( \rho_\beta \vec{j}_\alpha \cdot \nabla_\alpha + \rho_\alpha \vec{j}_\beta \cdot \nabla_\beta \right) g_{\alpha \beta}
  + \sqrt{\rho_\alpha \rho_\beta} \sumint_{\gamma,\eta} S_{\alpha \gamma} \deltaXTildeDot_{\gamma \eta} S_{\beta \eta}
  = \rho_\alpha \rho_\beta \sumint_{\gamma} Y_{\gamma \alpha \beta} \nabla_\gamma \cdot \vec{j}_\gamma\,.
\end{align}
\end{widetext}
We split the time dependence of $g_{\alpha \beta}$ into fluctuations of $\deltarho_\alpha$ and $\deltau_{\alpha \beta}$
as the hypernetted-chain equations give a relation between these three quantities,
$\dot{g}_{\alpha \beta} = \left( \partial_\rho + \partial_u \right) g_{\alpha \beta}$.
We use the OZE to determine the functional derivatives,
\begin{align}
  \partial_\rho g_{\alpha \beta}
  &= \sumint_{\gamma} \frac{\delta g_{\alpha \beta}}{\delta \rho_\gamma} \deltarho_\gamma
   = \sumint_{\gamma} Y_{\gamma \alpha \beta} \deltarhoDot_\gamma
  \\
  \partial_u \tilde{g}_{\alpha \beta}
  &= \sumint_{\gamma,\eta}\! \frac{\delta \tilde{g}_{\alpha \beta}}{\delta u_{\gamma \eta}} \deltauDot_{\gamma \eta}
   = \sumint_{\gamma,\eta}\! S_{\alpha \gamma} \left( \partial_u \delta \widetilde{X}_{\gamma \eta} \right) S_{\beta \eta}\,.
\end{align}
In the ULA we further have $\partial_u \deltaXTilde_{\alpha \beta} \approx \deltaXTildeDot_{\alpha \beta}$.

We define the one-body Hamiltonian-like operator
\begin{align}
  \hamiltonian_\alpha \equiv - \frac{\hbar^2}{2 m_\alpha} \frac{1}{\sqrt{\rho_\alpha}} \nabla_\alpha \cdot \rho_\alpha \nabla_\alpha \frac{1}{\sqrt{\rho_\alpha}}
\end{align}
and the Bijl-Feynman (BF) states $\psi_{n,\alpha}$ as the eigenvectors of the generalized eigenvalue problem
\begin{align}
  \label{eq:feynmanevp}
  \hamiltonian_\alpha \psi_{n,\alpha}
  = \varepsilon_n \sumint_{\beta} S_{\alpha \beta} \psi_{n,\beta}
  \text{.}
\end{align}
The BF states $\psi_{n,\alpha}$ are not orthonormal.  Therefore we define the adjoint states
\begin{align}
  \phi_{n,\alpha} \equiv \sumint_{\beta} S_{\alpha \beta} \psi_{n,\beta}
\end{align}
such that the following orthonormality relation is fulfilled:
\begin{align}
  \sumint_{\alpha} \phi^*_{n,\alpha} \psi_{m,\alpha} = \delta_{nm}
\end{align}
We introduce the abbreviation
\begin{align}
  \tilde{\zeta}_{n,\alpha} \equiv \phi_{n,\alpha} - \psi_{n,\alpha}
  \text{.}
\end{align}
With the expansions in terms of the BF states,
\begin{align}
  \deltarhoTilde_\alpha &= \sum\limits_{m} r_m(t) \phi_{m,\alpha} \label{eq:deltarhoTilde}\\
  \deltaXTilde_{\alpha \beta} &= \sum\limits_{m, n} X_{mn}(t) \psi_{m,\alpha} \psi_{n,\alpha} \\
  \potentialTilde_\alpha &= \sum\limits_{m} u_m(t) \psi_{m,\alpha}\,,
\end{align}
and the definitions
\begin{align}
  Z_{\alpha,mn} 
  &\equiv \frac{1}{\sqrt{\rho_\alpha}} \sumint_{\beta,\gamma} \phi_{m,\beta} \phi_{n,\gamma} \widetilde{X}_{\alpha \beta \gamma} \\
  \vec{W}_{\alpha,mn}
  &\equiv \phi_{m,\alpha} \nabla_\alpha \zeta_{n,\alpha}
  + \phi_{n,\alpha} \nabla_\alpha \zeta_{m,\alpha}
  + \sqrt{\rho_\alpha} \nabla_\alpha Z_{\alpha,mn} \\
  V_{pq,n}
  &\equiv \sumint_{\alpha} \frac{\hbar^2}{2 m_\alpha} \frac{\psi^\ast_{n,\alpha}}{\sqrt{\rho_\alpha}} \nabla_\alpha \cdot \sqrt{\rho_\alpha} \vec{W}_{\alpha,mn}
\end{align}
the EOMs \eqref{eq:eom1} and \eqref{eq:eom2alternate} can be written as
\begin{widetext}
\begin{align}
  \label{eq:eom1feynmanspace}
  i \hbar \dot{r}_m(t) - \varepsilon_m r_m(t) - \frac{1}{2} \sum\limits_{p,q} X_{pq}(t) V_{pq,n} = 2 u_m(t) \\
  \label{eq:eom2feynmanspace}
  i \hbar \dot{X}_{mn}(t) - \left( \varepsilon_m + \varepsilon_n \right) X_{mn}(t)
  - \sum\limits_{p,q} X_{pq}(t) \sumint_{\alpha} \frac{\hbar^2}{4 m_\alpha} \vec{W}_{\alpha,mn}^\ast \cdot \vec{W}_{\alpha,pq}
  &= \sum\limits_{p} r_p(t) V_{mn,p}^\ast
\end{align}
\end{widetext}
We solve these equations in frequency space and use an adiabatic Fourier transform,
\begin{align}
  \label{eq:adiabaticft}
  f(t) = \lim\limits_{\eta \rightarrow 0} e^{\eta t} \int \frac{\diff{\omega}}{2 \pi} e^{-i \omega t} f(\omega)\,,
\end{align}
for $r_m$, $X_{mn}$ and $u_m$.  Since the EOMs are linear, the Fourier transform simply results
in the substitutions $f(t) \rightarrow f(\omega)$ and $\dot{f}(t) \rightarrow (\eta - i \omega) f(\omega)$.
With the definition
\begin{align}
  A_{mnpq}(\omega) &\equiv \left( \hbar\omega - \varepsilon_m - \varepsilon_n + i \eta \right) \delta_{mp} \delta_{nq} \nonumber
  \\
  &\quad - \sumint_{\alpha} \frac{\hbar^2}{4 m_\alpha} \vec{W}_{\alpha,mn}^\ast \cdot \vec{W}_{\alpha,pq} 
  \label{eq:Aapprox}
  \\
  &\approx \left( \hbar\omega - \varepsilon_m - \varepsilon_n + i \eta \right) \delta_{mp} \delta_{nq}
\end{align}
we can then write \eqref{eq:eom2feynmanspace} in the frequency space as
\begin{align}
  \sum\limits_{p,q} A_{mnpq}(\omega) X_{pq}(\omega) = \sum\limits_{o} r_o(\omega) V_{mn,o}^\ast
  \text{.}
\end{align}
Note that we only use the diagonal part of $A_{mnpq}(\omega)$, \eqref{eq:Aapprox}. Then this relation can easily be inverted,
\begin{align}
  \label{eq:Xmn}
  X_{mn}(\omega) = \sum\limits_{o} \frac{r_o(\omega) V_{mn,o}^\ast}{\hbar\omega - \varepsilon_m - \varepsilon_n + i \eta }
  \text{.}
\end{align}
We define the self energy
\begin{align}
  \Sigma_{mn}(\omega) \equiv \frac{1}{2} \sum\limits_{p,q} \frac{V_{pq,m} V_{pq,n}^\ast}{\hbar\omega - \varepsilon_p - \varepsilon_q + i \eta}
\end{align}
and bring \eqref{eq:eom1feynmanspace} into frequency space using \eqref{eq:adiabaticft}, insert $X_{mn}(\omega)$ from \eqref{eq:Xmn} and use $\Sigma_{mn}(\omega)$ to get
\begin{align}
  \sum\limits_{n} \left\lbrace 
    \left( \hbar\omega - \varepsilon_m + i \eta \right) \delta_{mn} - \Sigma_{mn}(\omega) 
    \vphantom{x^{x^{x}}}
  \right\rbrace r_n(\omega) = 2 u_m(\omega)
  \text{.}
\end{align}
We introduce the Green function matrix
\begin{align}
  G_{mn}^{-1}(\omega)
  \equiv \left( \hbar\omega - \varepsilon_m + i \eta \right) \delta_{mn} - \Sigma_{mn}(\omega)
\label{eq:green1}
\end{align}
to finally solve the EOMs for the expansion coefficients of $\deltarhoTilde_\alpha$, eq.~(\ref{eq:deltarhoTilde}),
\begin{align}
  r_m(\omega) = 2 \sum\limits_{n} G_{mn}(\omega) u_n(\omega) \,\text{.}
\end{align}
This is the linear response relation between perturbations $u_m$ and fluctuations $r_m$, in the basis of the BF states.
In order to go back to coordinate space,
we first introduce the physical density fluctuations in frequency space
\begin{align}
  f_\alpha(\omega)
  \equiv \int\! \diff{t}\, e^{i \omega t} \Re \deltarhoTilde_\alpha(t)
\end{align}
which leads to the linear reponse of the physical density
\begin{align}
  f_\alpha(\omega)
  = \sumint_{\beta} \sum\limits_{m,n} \big(& \phi_{m,\alpha} \phi_{n,\beta}^\ast G_{mn}(\omega) \nonumber\\
  +& \phi_{m,\alpha}^\ast \phi_{n,\beta} G_{mn}^\ast(-\omega) \big) \potentialTilde_\beta(\omega)
  \,\text{.}
\label{eq:f}
\end{align}
The linear density-density response matrix $\chi_{\alpha \beta}(\omega)$ is defined via the linear relation
\begin{align}
  f_\alpha(\omega) = \sumint_{\beta} \chi_{\alpha \beta}(\omega) \potentialTilde_\beta(\omega)
\label{eq:fchiv}
\end{align}
and therefore we can immediately read off the result for $\chi_{\alpha \beta}(\omega)$ from eq.~(\ref{eq:f}),
\begin{align}
  \chi_{\alpha \beta}(\omega)
  = \sum\limits_{m,n} \left( \phi_{m,\alpha} \phi_{n,\beta}^\ast G_{mn}(\omega) + \phi_{m,\alpha}^\ast \phi_{n,\beta} G_{mn}^\ast(-\omega) \right)
  \text{.}
\end{align}
For the numerical evaluation of the response matrix one last thing has to be approximated: the triplet correlations $\widetilde{X}_{\alpha \beta \gamma}$. This is done as in Ref.~\onlinecite{Clements93}, generalized here for the multi-component case.
We write the triplet correlations as an expansion in terms of BF states.
\begin{align}
  \widetilde{X}_{\alpha \beta \gamma}
  &= \sum\limits_{m,n,o} \psi_{m,\alpha} \psi_{n,\beta}^\ast \psi_{o,\gamma}^\ast \frac{V_{mno}}{ \varepsilon_m + \varepsilon_n + \varepsilon_o }\,.
\end{align}
Using the definitions
\begin{align}
  \Xi_{\alpha,mno} &\equiv \phi_{m,\alpha}^\ast \left( \nabla_\alpha \zeta_{n,\alpha} \right) \cdot \left( \nabla_\alpha \zeta_{o,\alpha} \right) \\
  \Omega_{\alpha,mno} &\equiv \phi_{m,\alpha} \left( \nabla_\alpha \zeta^\ast_{n,\alpha} \right) \cdot \left( \nabla_\alpha \zeta_{o,\alpha} \right)
\end{align}
we can approximate the coefficients\cite{Clements93}
\begin{align}
  V_{mno} &\approx -\sumint_{\alpha} \frac{\hbar^2}{2 m_\alpha} \sqrt{\rho_\alpha} \left( \Xi_{\alpha,mno} + \Omega_{\alpha,nmo} + \Omega_{\alpha,omn} \right)
  \,\text{.}
\end{align}

\section{Application to homogeneous systems}

In this work we are interested in translationally invariant layers
of two-dimensional dipolar Bose gases, where the different layers correspond
to different components.  In the homogeneous limit the BF states are plain waves,
\begin{align}
  \psi_{n,\alpha}(\rv_\alpha) 
  \longrightarrow
  e^{i \kv \cdot \rv_\alpha} \psi_{n,\alpha}(\kv)
  \,\text{.}
\label{eq:BFhom}
\end{align}
Hence the quantum number $n$ from the previous section becomes $(n,\kv)$ and $\sum_n$ becomes $\sum_n\int\! {\diff{k}\over(2\pi)^2}$. $\alpha=1,\dots,L$ is the layer index, and $n=1,\dots,L$ labels the $L$ BF modes of this
$L$-component Bose gas, in ascending order of their energy.  If the layers are arranged symmetrically
(layers $i$ and $L-i+1$ are identical),
$\psi_{n,\alpha}(\kv)$ are vectors with index $\alpha$ which are alternatingly symmetric/antisymmetric with the lowest
BF state being symmetric.  The BF
states (\ref{eq:BFhom}) are solutions of eq.~(\ref{eq:feynmanevp}), which leads to the generalized eigenvalue
equation for $\psi_{n,\alpha}(\kv)$ (assuming particles of the same mass in all layers)
\begin{align}
  {\hbar^2 k^2\over 2m} \psi_{n,\alpha}(\kv) = \varepsilon_n(k) \sum_\beta S_{\alpha\beta}(\kv)\psi_{n,\beta}(\kv)\,,
\label{eq:feynmanhom}
\end{align}
with the structure function matrix
\begin{align}
  S_{\alpha\beta}(\kv) = \delta_{\alpha\beta} + \sqrt{\rho_\alpha\rho_\beta}\int\! \diff{\rv}\, e^{-i\kv\rv} h_{\alpha\beta}(\rv)\,.
\label{eq:Shom}
\end{align}

In the homogeneous limit, the self energy becomes
\begin{align}
  \Sigma_{mn}(\kv_m,\kv_n,\omega) = \delta(\kv_m-\kv_n) \bar\Sigma_{mn}(\kv_m,\omega)
\end{align}
with the matrix
\begin{widetext}
\begin{align}
  \bar\Sigma_{mn}(\kv_m,\omega) =
  \frac{1}{2} \sum\limits_{p,q} \int {\diff{\kv}_p\over(2\pi)^2}
  \frac{\bar V_{pq,m}(\kv_p,\kv_m) \bar V_{pq,n}^\ast(\kv_p,\kv_m)}{\hbar \omega - \varepsilon_p(\kv_p) - \varepsilon_q(\kv_m-\kv_p) + i \eta}\,,
\label{eq:sigmahom}
\end{align}
\end{widetext}
where $\bar V_{pq,m}(\kv_p,\kv_m)$ is defined by factoring out the $\delta$ distribution,
\begin{align}
  V_{pq,m}(\kv_p,\kv_q,\kv_m) = \bar V_{pq,m}(\kv_p,\kv_m) \delta(\kv_n-\kv_p-\kv_q)\,.
\end{align}
For a symmetric arrangement of layers, $\bar V_{pq,m}(\kv_p,\kv_m)$ vanishes if $p+q+m$ is odd.
The Green function (\ref{eq:green1}) becomes
\begin{align}
  G_{mn}(\kv_m,\kv_n,\omega) = \delta(\kv_m-\kv_n) \bar G_{mn}(\kv_m,\omega)
\end{align}
with the matrix
\begin{align}
  \bar G^{-1}_{mn}(\kv_m,\omega) =
  &\big(\hbar \omega - \varepsilon_m(\kv_m) + i\eta\big)\delta_{mn} \nonumber\\
  & - \bar\Sigma_{mn}(\kv_m,\omega)\,.
\end{align}
The response function in momentum space becomes the matrix
\begin{align}
  \chi_{\alpha \beta}(\kv,\omega) = \sqrt{\rho_\alpha\rho_\beta}\sum_{m,n} &\phi_{n,\alpha}(\kv)\phi_{m,\beta}^*(\kv)
                                  \big[ \bar G_{mn}(\kv,\omega)\nonumber\\
  &+ \bar G_{mn}^*(-\kv,-\omega)\big]\,,
\label{eq:chihom}
\end{align}
where we used $\phi^*_{n,\alpha}(\kv)=\phi_{n,\alpha}(-\kv)$.  Using eq.~(\ref{eq:f}), we see that
the matrix $\chi_{\alpha \beta}(\kv,\omega)$ provides the linear relation between perturbations $\potentialTilde_\beta(\kv,\omega)$ of layers $\beta$
(with frequency $\omega$ and wave number $\qk$) and the density response $\Delta\rho_\alpha=\sqrt{\rho_\alpha}f_\alpha$ of layers $\alpha$,
\begin{align}
\Delta\rho_\alpha(\kv,\omega) =\sum_\beta \chi_{\alpha \beta}(\kv,\omega)\, V_\beta(\kv,\omega)\,.
\label{eq:Deltarho}
\end{align}
Ostensibly, due to the coupling between different layers, a perturbation of layer $\beta$ leads to
a density response in all layers $\alpha$.

\end{appendix}

\bibliography{bec,papers,my}
\end{document}